%% file: AGN12_May2018.tex
\documentclass[twocolumn]{aastex6}

\usepackage{epsfig}
\usepackage{amsmath}
\usepackage{xspace}
\usepackage{graphicx}
\usepackage{natbib}
\usepackage{bm}

\newcommand{\kms}{\ifmmode {\rm km~s}^{-1} \else km~s$^{-1}$\fi}
\newcommand{\ergs}{\ifmmode {\rm erg~ s}^{-1} \else erg~s$^{-1}$\fi}
\newcommand{\ergscm}{\ifmmode {\rm erg~s}^{-1} \else erg~s$^{-1}$ cm$^{-2}$\fi}
\newcommand{\Msun}{\ifmmode M_{\odot} \else $M_{\odot}$\fi }
\newcommand{\Lsun}{\ifmmode {\rm L}_{\odot} \else L$_{\odot}$\fi}
\newcommand{\qo}{\ifmmode q_{\rm o} \else $q_{\rm o}$\fi}
\newcommand{\Ho}{\ifmmode H_{\rm o} \else $H_{\rm o}$\fi}
\newcommand{\ho}{\ifmmode h_{\rm o} \else $h_{\rm o}$\fi}

\newcommand{\vFWHM}{\ifmmode v_{\mbox{\tiny FWHM}} \else
                    $v_{\mbox{\tiny FWHM}}$\fi}
\newcommand{\CCF}{\ifmmode F_{\it CCF} \else $F_{\it CCF}$\fi}
\newcommand{\ACF}{\ifmmode F_{\it ACF} \else $F_{\it ACF}$\fi}
\newcommand{\Halpha}{\ifmmode {\rm H}\alpha \else H$\alpha$\fi}
\newcommand{\Hbeta}{\ifmmode {\rm H}\beta \else H$\beta$\fi}
\newcommand{\Hgamma}{\ifmmode {\rm H}\gamma \else H$\gamma$\fi}
\newcommand{\Hdelta}{\ifmmode {\rm H}\delta \else H$\delta$\fi}
\newcommand{\Lya}{\ifmmode {\rm Ly}\alpha \else Ly$\alpha$\fi}
\newcommand{\Lyb}{\ifmmode {\rm Ly}\beta \else Ly$\beta$\fi}
\newcommand{\HeI}{\ifmmode {\rm He}\,{\sc i}\,\lambda5876 \else 
	          He\,{\sc i}\,$\lambda5876$\fi}
\newcommand{\HeII}{\ifmmode {\rm He}\,{\sc ii}\,\lambda4686 \else 
	           He\,{\sc ii}\,$\lambda4686$\fi}

\newcommand{\heii}{\ifmmode \makebox{{\rm He}\,{\sc ii}} \else He\,{\sc ii}\fi}

\newcommand{\feii}{Fe\,{\sc ii}}

\newcommand{\ciii}{\ifmmode {\rm C}\,{\sc iii} \else C\,{\sc iii}\fi}

\newcommand{\oiii}{O\,{\sc iii}}

\def\fake2{\hphantom{3}}

\shorttitle{Velocity-Resolved reverberation mapping of five bright seyfert 1 galaxies}
\shortauthors{De Rosa et al.}

\begin{document}

\title{Velocity-Resolved reverberation mapping of five bright seyfert 1 galaxies}

\author{G.~De~Rosa\altaffilmark{1,2,3}, 
 M.M.~Fausnaugh\altaffilmark{1}, 
 C.J.~Grier\altaffilmark{1,4,5}, 
 B.M.~Peterson\altaffilmark{1,2,3}, 
 K.D.~Denney\altaffilmark{1,2,6,7}, 
 Keith~Horne\altaffilmark{8}, 
 M.C.~Bentz\altaffilmark{9}, 
 S.~Ciroi\altaffilmark{10}, 
 E.~Dalla~Bont\`a\altaffilmark{10,11}, 
 M.D.~Joner\altaffilmark{12}, 
 S.~Kaspi\altaffilmark{13,14}
% ...
}

\altaffiltext{1}{Department of Astronomy, The Ohio State University, 140 W 18th Ave, Columbus, OH 43210, USA} 
\altaffiltext{2}{Center for Cosmology \& AstroParticle Physics, The Ohio State University, 191 West Woodruff Ave, Columbus, OH 43210, USA}
\altaffiltext{3}{Space Telescope Science Institute, 3700 San Martin Drive, Baltimore, MD 21218, USA} 
\altaffiltext{4}{Department of Astronomy and Astrophysics, Eberly College of Science, The Pennsylvania State University, 525 Davey Laboratory, University Park, PA 16802, USA}
\altaffiltext{5}{Institute for Gravitation and the Cosmos, The Pennsylvania State University, University Park, PA 16802, USA}
\altaffiltext{6}{NSF Postdoctoral Research Fellow} 
\altaffiltext{7}{Illumination Works, LLC, 5650 Blazer Parkway, Dublin, OH 43017, USA}
\altaffiltext{8}{SUPA Physics and Astronomy, University of St.\ Andrews, Fife, KY16 9SS Scotland, UK}
\altaffiltext{9}{Department of Physics and Astronomy, Georgia State University, 25 Park Place, Suite 605, Atlanta, GA 30303, USA}
\altaffiltext{10}{Dipartimento di Fisica e Astronomia ``G. Galilei,'' Universit\`{a} di Padova, Vicolo dell'Osservatorio 3, I-35122 Padova, Italy}
\altaffiltext{11}{INAF-Osservatorio Astronomico di Padova, Vicolo dell'Osservatorio 5 I-35122, Padova, Italy}
\altaffiltext{12}{Department of Physics and Astronomy, N283 ESC, Brigham Young University, Provo, UT 84602, USA}
\altaffiltext{13}{Wise Observatory and School of Physics and Astronomy, Raymond and Beverly Sackler Faculty of Exact Sciences, Tel Aviv University, Tel Aviv 69978, Israel}
\altaffiltext{14}{Physics Department, Technion, Haifa 32000, Israel}
% ...

\begin{abstract}
  We present the first results from a reverberation-mapping campaign
  undertaken during the first half of 2012, with additional data on
  one AGN (NGC 3227) from a 2014 campaign. Our main goals
  are (1) to determine the black hole masses from continuum-H$\beta$
  reverberation signatures, and (2) to look for velocity-dependent
  time delays that might be indicators of the gross kinematics of the
  broad-line region. We successfully measure H$\beta$ time delays and
  black hole masses for five AGNs, four of which have previous
  reverberation mass measurements.  The values measured here are in
  agreement with earlier estimates, though there is some intrinsic
  scatter beyond the formal measurement errors. We observe 
  velocity dependent H$\beta$ lags in each case, and 
  find that the patterns have changed in the intervening five years for three AGNs that were also observed in 2007.
\end{abstract}


\section{INTRODUCTION}
\label{section:intro}

Variability of the broad emission-line fluxes and profiles is commonly
seen in the spectra of Type 1 active galactic nuclei (AGNs).  A number
of isolated cases of dramatic emission-line changes were reported
based on photographic spectrograms by the late 1960s and early 1970s
(e.g., \citealt{Andrillat68,Pastoriza70}; see the reviews by
\citealt{Pronik80} and by \citealt{Collin80}).  Additional and more
convincing instances of emission-line changes were found on
surprisingly short timescales with the advent of linear detectors for
spectrometers on ground-based telescopes
\citep[e.g.,][]{Tohline76,Boksenberg77,Foltz81,Kollatschny81,Schulz81,
  Peterson82,Antonucci83} and in the UV with the {\em International
  Ultraviolet Explorer} \citep[e.g.,][]{Ulrich84}. The interested
reader is referred to \cite{Peterson88} for a review of the early
studies of emission-line variability in AGNs.

That correlated variability of continuum and emission-line fluxes
could be used to probe the structure of the broad-line region (BLR) in
active galactic nuclei (AGNs) was recognized in the first decade of
quasar research \citep{Bahcall72}. The concept was refined in the
early 1980s and has been known since as ``reverberation mapping''
\citep{Blandford82} because the emission lines ``reverberate'' in
response to continuum variations.  Reverberation mapping has since
become a standard tool for studying the structure and dynamics of the
BLR \citep{Peterson93,Peterson14}. Many programs were undertaken in
the 1990s, largely enabled by the proliferation of high-quality
detectors on small to medium-sized telescopes where groups of
observers could obtain enough telescope time for long-term monitoring
campaigns.

In its simplest form, reverberation mapping is used to measure the
mean response time $\tau$ of emission lines to continuum variations, and this
is interpreted as the light travel-time across the BLR radius $R = c\tau$.
By combining the measured time delay, or lag, between continuum
and emission-line flux variations with some suitable measure of the
emission-line width $\Delta V$, it is possible to estimate the
mass of the super massive black hole that is the central
engine of the AGN. The mass is usually expressed as
\begin{equation}
\label{eq:mass}
M_{\rm BH} = f \left( \frac{c \tau \Delta V^2}{G} \right),
\end{equation}
where $G$ is the gravitational constant and $f$ is an unknown scaling
constant. The quantity in parentheses is often referred to as the
``virial product'' (VP), which has units of mass and contains only the
two observables ($\tau$ and $\Delta V$) and physical constants. All
complicating factors, such as the inclination of the system or the
effects of anisotropic line emission, are subsumed into the constant
$f$.  Thus, $f$ is expected to be different for every individual AGN,
but should be approximately constant for every emission line in a
given AGN assuming similar geometries and dynamics of the
line-emitting gas.  In every case where the lags from multiple
emission lines can be measured in a single object, it is found that $\tau \propto
\Delta V^{-2}$ as expected from Equation (\ref{eq:mass}) suggesting
that this is the case
\citep{PW99,PW00,Kollatschny03,Peterson04,Bentz10a}.
It is worth reminding the reader that the scale factor $f$
depends on which parameter is used to characterize the emission-line width, as we discuss in \S{\ref{section:linewidths}}.

The scaling
factor $f$ can be determined for an AGN if there is an independent
measurement of the black hole mass. Unfortunately, there are few cases
where the black hole radius of influence is large enough that either
stellar or gas dynamical modeling can also be used. At the present
time, there are stellar dynamical masses \citep{Davies06,Onken14} and
gas dynamical masses \citep{Hicks08} for NGC 3227 and NGC 4151; these
are useful for comparison purposes, but it would be unwise to attempt
to calibrate the entire reverberation-based mass scale on only two
objects.  Instead, one of the well-known correlations between
central black hole mass and properties of the host galaxies
can be used. The first of these
to be used to calibrate the AGN black hole mass scale was the 
correlation between black hole mass and host bulge luminosity
\citep{Magorrian98,Laor98,McLure01,McLure02,Haring04}. More recently,
calibration of the AGN black hole mass scale has been based on the
strong correlation
between the black hole mass and the velocity dispersion of host-galaxy
bulge, the $M_{\rm BH}$--$\sigma_*$ relationship, which applies to
both quiescent \citep{Ferrarese00,Gebhardt00a,
  Tremaine02,Gultekin09,McConnell11,McConnell13} and active galaxies
\citep{Gebhardt00b,Ferrarese01,Nelson04,
  Onken04,Dasyra07,Woo10,Graham11,Park12,Grier13b,Woo15, Batiste17}.  With
$\sigma_*$ measurements now available for $\sim 30$ AGNs from the
reverberation mapping database \citep{Woo15}, an ensemble average
$\langle f \rangle = 4.47 \pm 1.25$ can be computed by comparing the
predicted masses from the $M_{\rm BH}$--$\sigma_*$ relationship with
the observed virial products, using the line dispersion to characterize
the line width.  Using this prescription, black hole
masses have been measured for $\sim60$ AGNs using reverberation
mapping (see \citealt{Bentz15} for an up-to-date compilation).

An important result from reverberation mapping is the observed ``$R$--$L$''
relationship between the size of the BLR and the AGN luminosity 
\citep{WPM99,Kaspi00,Kaspi05,Bentz06b,Bentz09,Bentz13}. This
$R$--$L$ relationship allows us to bypass resource-intensive reverberation
mapping by using the luminosity to infer the BLR radius. By combining
the estimate of the BLR radius with the emission-line width, we can
apply Equation (\ref{eq:mass}) to estimate the black hole mass  \citep[see][for a review on single epoch $M_{\rm BH}$ estimates] {Vestergaard2011}.

The frontier of reverberation mapping is determination of
the kinematics and geometry of the BLR by examination of the emission-line
response as a function of line-of-sight velocity. The ultimate goal is to either model
the BLR geometry and kinematics directly \citep[e.g.,][]{Pancoast12,Pancoast14,Waters16} or 
to recover velocity--delay maps and model the BLR indirectly
\citep[e.g.,][]{Bentz10b,Grier13a}. 
Observational results are only now beginning to
appear as the technical requirements for detailed reverberation mapping
are quite demanding \citep{Horne04}.

Over the last decade, we have undertaken a new series of reverberation programs
with several specific goals in mind:
\begin{enumerate}
\item To increase the number of AGNs for which reverberation lags are
  measured for the \Hbeta\ emission line.  Additional data can better
  constrain the $R$--$L$ relationship \citep{Bentz13} and the AGN
  $M_{\rm BH}$--$\sigma_*$ relationship that underlies the
  reverberation-based black hole mass calibration scale
  \citep{Grier13b}.
\item To improve upon previous reverberation results. Our reanalysis
  of nearly all the reverberation data that existed a decade ago
  revealed that many of the sources would benefit from a higher
  sampling rate \citep{Peterson04}.
\item To obtain higher quality, higher time-resolution spectra that
  would enable recovery of velocity--delay maps
  \citep[e.g.,][]{Grier13a}.
\end{enumerate}
These programs were designed to meet the criteria described by
\cite{Horne04} to enable recovery of velocity--delay maps. They were carried out at MDM and partner observatories
in 2005 \citep{Bentz06a,Denney06,Bentz07}, 2007
\citep{Denney09a,Denney09b,Denney10}, and 2010
\citep{Grier12a,Grier12b,Grier13a}.  In addition to these ground-based
programs, we carried out an intensive multiwavelength campaign on
NGC~5548 known as the AGN Space Telescope and Optical Reverberation
Mapping (AGN STORM) project
\citep{DeRosa15,Edelson15,Fausnaugh16,Goad16,Pei17,Starkey17,Mathur17} and a
concurrent optical monitoring program on additional AGNs
\citep{Fausnaugh17b}. The amount of AGN reverberation
data has increased dramatically over the last few years, with several other groups carrying out campaigns similar to ours
\citep{Bentz09b,Bentz10b,Barth11a,Barth11b,Barth13,Bentz14,Du14,Pei14,
  Wang14,Barth15,Du15, Bentz16a,Bentz16b, Du16}.
On-going large multi-object reverberation-mapping campaigns
\citep{King15,Shen15} are expected to significantly increase the 
number of reverberation-mapped AGNs, as well as increase redshift and luminosity ranges of the sample, especially for emission lines other than the Balmer series.

Here
we report results from a campaign undertaken in early 2012. We also
include additional results on NGC\,3227 from 2014. We describe the observations and data analysis
in \S{2}. Our time-series analysis is presented in \S{3} and
our black hole mass measurement is explained in \S{4}.
We briefly discuss and summarize our results in \S{5}. When needed, we adopt a
cosmological model with $\Omega_{m} = 0.30$, $\Omega_{\Lambda} = 0.70$, and
$H_0 = 70$ km sec$^{-1}$ Mpc$^{-1}$.

\section{OBSERVATIONS \& DATA ANALYSIS}
\label{section:obs}

\subsection{Target Selection}
\label{section:targets}

The primary objective of this campaign is to determine the kinematics
and structure of the BLR in a few well-studied bright AGNs. In
particular, we are re-examining NGC\,3227, NGC\,3516, and NGC\,5548
from \cite{Denney09a}, for which cross-correlation of individual
velocity-bins suggested gross kinematics of outflow, infall, and
rotation/virialization, respectively.  As we discuss here and
elsewhere, these results need to be checked and more thoroughly
characterized.  We also included in our observing program NGC\,4151,
for which the best reverberation data are from a weather-abbreviated campaign in 2005
\citep{Bentz06a}. In addition to these primary targets, we added
 a few sources that could only be observed for part of our
campaign on account of their location in the sky. Sources included in
the 2012 campaign are Mrk 374, Mrk 382, Mrk 478, Mrk 618, and Mrk
704. Because of the shorter monitoring period, the failure rate for
these secondary sources was high, with only Mrk 704 yielding data
useful for reverberation purposes.

The properties of the sources studied in this paper are summarised in Table
1. 
Both NGC~3227 and NGC 4151 are too close for redshift-based distances
to be reliable.  NGC 3227 is interacting with an elliptical companion,
NGC 3226, which has a surface-brightness fluctuation distance of
23.5\,Mpc \citep{Tonry01}. We therefore adopt this as the distance to
NGC 3227. In the case of NGC 4151, we are currently working on a
Cepheid-based distance, but here we use the distance of 13.9\,Mpc
adopted by \citet{Onken14} in their recent stellar dynamical study
(although this distance is derived from Hubble's law).

\subsection{Observations}
\label{sec:obsall}

\subsubsection{Spectroscopy}
\label{section:specobs}

The principal data source for both the 2012 and 2014 campaigns was the 
Boller and Chivens CCD spectrograph
on the MDM Observatory 1.3-m McGraw-Hill telescope on Kitt Peak.
The 2012 campaign ran from the beginning of 2012 January through the end of 2012 April.
We used a 350 mm$^{-1}$ grating to obtain a
dispersion of 1.33\,\AA\,pixel$^{-1}$. We set the grating for a
central wavelength of 5150\,\AA, which resulted in spectral coverage
over the range 4400\,\AA \ to 5850\,\AA. The slit was oriented
north--south (position angle ${\rm PA} = 0^{\rm o}$) with a 
projected width of $5\farcs0$ that results in
a spectral resolution of 7.9\,\AA. We used an extraction
window of $12\farcs0$ along the slit.

The 2012 campaign also included spectroscopic observations obtained at
the Asiago Astrophysical Observatory of Padova University with the
1.22-m {\em Galileo} telescope and the Cassegrain Boller
$\&$ Chivens spectrograph. We used a 300 mm$^{-1}$ grating in first
order combined with a $5\farcs0 \times 7\farcm75$ long slit oriented
at ${\rm PA} = 90^{\rm o}$.
The spectral range between about 3200 \AA\ and 8000 \AA\ was covered
with a dispersion of 2.3 \AA\ pixel$^{-1}$. The spatial scale is 1
arcsec pixel$^{-1}$, the  resulting resolution is 10.5\,\AA. 
We used an extraction window of $12\farcs0$. 

The Crimean Astrophysical Observatory (CrAO) provided spectra from 
the Nasmith spectrograph and SPEC-10 CCD on the 2.6 m Shajn telescope. 
We used a $3\farcs0$ slit at a position angle of $90^{\rm o}$, and an extraction 
window of $11\farcs0$. The CrAO data cover wavelengths from 3900\,\AA \ to 
6100\,\AA, with a dispersion of 1.85 \AA\ pixel$^{-1}$. The smaller size of the 
slit for the CrAO configuration compared to the MDM and Asiago observations
introduces a different amount of host galaxy light in the extracted spectra. 
However, the galaxy flux is not variable in time and we correct for this in the 
final light curves (\ref{section:lightcurves}).

Finally, the 2.3-m telescope at Wyoming Infrared Observatory
(WIRO) and the WIRO Long Slit Spectrograph contributed a small number of 
observations, to fill in planned gaps during the monitoring campaign.
We used a 900 mm$^{-1}$ grating, resulting in a
$\sim$1\,\AA\ pixel$^{-1}$ dispersion between 4400\,\AA\ and
5600\,\AA.  A $5\farcs0$ slit aligned at ${\rm PA}=0^{\rm o}$ was used
with a $12\farcs0$ extraction window.

\subsubsection{Imaging}
\label{section:imagedata}

We supplemented our spectroscopic continuum light curves 
with photometric observations. Observations in 2012 were obtained with 
the 0.5-m Centurian 18 at Wise Observatory (WC18, \citealt{Brosch08})
and with the 0.9-m at West Mountain Observatory (WMO).
WC18 uses a STL-6303E CCD with
a pixel scale of $1\farcs47$ and a $75\arcmin \times 50\arcmin$ field of view, and WMO
uses a Finger Lakes PL-3041-UV CCD with a pixel scale of
$0\farcs61$ and a field of view of $21\arcmin \times 21\arcmin$.
We also used data from the All-Sky Automated Survey for SuperNovae 
(ASAS-SN, \citealt{Shappee14}). These data are from the first unit of 
ASAS-SN, Brutus, which consisted in 2012 of two 14 cm aperture Nikon 
telephoto lenses on a single mount in the Faulkes Telescope North \citep{Brown2013} 
enclosure on Mount Haleakala, Hawaii. ASAS-SN detectors are FLI ProLine CCD 
cameras, each with a Fairchild Imaging 2k$\times$2k thinned CCD, 
a $4.47\times4.47$ degree field of view, and a $7\farcs8$ pixel scale.

In addition, 
the 2014 campaign included
imaging from Crimean Astrophysical Observatory (CrAO), Fountainwood Observatory (FWO), and the Las Cumbres Observatory (LCO, \citealt{Brown2013}).
The CrAO images are from the 0.7-m telescope equipped with an AP7p CCD 
with a pixel scale of $1\farcs76$ and a field of view of $15\arcmin \times
15\arcmin$. Observations from FWO were obtained with a 0.4-m telescope with an
SBIG 8300M CCD. The field of view is $19\arcmin \times 17\arcmin$
and the pixel scale is $0\farcs35$.
The LCO data were obtained using their world-wide network of 1-m telescopes 
in the Sloan $ugriz$ bands.

\subsection{Data Processing and Light Curves}
\label{section:dataproc}

The procedures we followed for reducing the data, producing calibrated
light curves, and assessing uncertainties are described in detail by
\cite{Fausnaugh17b}.  We provide a brief recapitulation here.

\subsubsection{Spectroscopy}
\label{section: speccal}

Two-dimensional spectra were reduced using standard {\tt IRAF} tasks
to deal with bias, flat field, sky subtraction, and wavelength
calibration.  An extraction window of $12\arcsec$ was used
throughout. Cosmic ray removal was done using {\tt LA Cosmic}
\citep{vanDokkum01}. Flux calibration relied on observations of
standard stars, usually Feige 34 and/or BD $+33^{\rm o}2642$
\citep{Oke90}.

We used the narrow [O\,{\sc iii}]\,$\lambda5007$ emission line as an
internal flux standard for both relative and absolute calibration.
While narrow emission lines have been found to vary on long timescales
\citep[years to decades, e.g.,][]{Peterson13}, they are effectively constant in flux on
BLR reverberation timescales (days to months).  We identified all the individual
spectra where the observer reported ``clear'' or ``photometric''
observing conditions. The [O\,{\sc iii}]\,$\lambda5007$ flux was
measured, and from these a mean and standard deviation was
computed. Outliers greater than $3\sigma$ from the mean were rejected 
and the mean and standard deviation were recomputed. The number of observations
used for the calibration is given in Column (2) of Table
2 
and the adopted [O\,{\sc iii}]\,$\lambda5007$ fluxes appear in Column
(3) of the same table. This provides the absolute flux calibration for
the spectrophotometric observations. We note that the [O\,{\sc
  iii}]\,$\lambda5007$ flux in NGC 5548 is in good agreement with the
preliminary measurement we presented earlier \citep{Peterson13}.

For each AGN, the spectra with the highest signal-to-noise ratios and
no obvious flaws are combined to form a reference spectrum, which is
scaled to have the adopted [O\,{\sc iii}]\,$\lambda5007$ flux.  We
then place all the individual spectra on the same relative flux by
scaling each spectrum to the adopted [O\,{\sc iii}]\,$\lambda5007$
flux.  This is done using a Monte Carlo Markov Chain (MCMC) code
called {\tt mapspec} \citep{Fausnaugh17a} that adjusts the flux,
wavelength shift, and resolution of each individual spectrum to match that
of the reference spectrum, as measured by the [O\,{\sc
  iii}]\,$\lambda5007$ emission line profiles.  This affords a factor
of several improvement over the long-used method of
\cite{vanGroningen92}, as assessed by the root-mean-square scatter of
the [O\,{\sc iii}]\,$\lambda5007$ flux across the full time series.

Once flux calibration is complete, we combine all $N$ spectra for each object to form
a weighted mean spectrum
\begin{equation}
\label{eq:meanspec}
\langle F (\lambda) \rangle = \frac{\sum_{i=1}^{N} F(\lambda, t_i)/\sigma^2(\lambda, t_i)}
{\sum_{i=1}^{N} 1/\sigma^2(\lambda, t_i)},
\end{equation}
where $F(\lambda, t_i)$ is the flux at epoch $t_i$ and $\sigma(\lambda, t_i)$ is the associated
uncertainty. We also form a root-mean-square (RMS) residual spectrum
\begin{equation}
\label{eq:RMSspec}
\sigma_{\rm rms}(\lambda) = \left\{ \frac{1}{N-1} \sum_{i=1}^{N} \left[ F(\lambda, t_i) - 
\langle F(\lambda) \rangle \right]^2 \right\}^{1/2}.
\end{equation}
The mean and RMS spectra for our sources are shown in Figures
\ref{fig:mrk704data} -- \ref{fig:n5548data}.  The RMS spectrum is of
special value in this context, since the constant components (e.g.,
host-galaxy starlight, narrow emission lines) vanish, isolating the
variable part of the spectrum. However, the total variability power
contains contributions not only from intrinsic variability, but also
from statistical fluctuations and/or measurement errors. We therefore
attempt to isolate the intrinsic variability by minimizing the
negative log-likelihood
\begin{eqnarray}
\label{eq:intrinsicRMSspec}
-2 \ln {\cal L} & = & \sum_{i=1}^{N} \frac{\left[ F(\lambda, t_i) - \hat{F}(\lambda) \right]^2}
{\sigma^2(\lambda, t_i) + \sigma_{\rm var}^2(\lambda)} \nonumber \\
& &+ \sum_{i=1}^{N} \ln \left[ \sigma^2(\lambda, t_1) + \sigma_{\rm var}^2(\lambda) \right],
\end{eqnarray}
where $\hat{F}(\lambda)$ is the optimal average weighted by
$\sigma^2(t_i) + \sigma^2_{\rm var}$ and $\sigma_{\rm var}(\lambda)$ is the intrinsic variability.
We solve simultaneously for $\hat{F}(\lambda)$ and $\sigma_{\rm var}(\lambda)$.
Our estimator of the intrinsic variability $\sigma_{\rm var}(\lambda)$
is also shown in Figures \ref{fig:mrk704data} -- \ref{fig:n5548data}.

\subsubsection{Imaging}
\label{section:imagecal}
 
Independent continuum light curves were generated for each bandpass
for each set of imaging data using the image subtraction software
package {\tt ISIS} \citep{Alard98,Alard00}.  We followed the
procedures as described by \cite{Shappee11}. First, we aligned the
images using {\tt Sexterp} \citep{Siverd12}.  We then created a
reference image with {\tt ISIS} for each AGN field by combining the
images with the best seeing and lowest background counts; typically we
used 5--15 images to construct the reference image. {\tt ISIS}
convolves the images of each AGN with a convolution kernel that is
allowed to vary across the field in order to transform all the images
to the same point-spread function (PSF) and background level. The
reference image was convolved to match each individual frame and {\tt
  ISIS} then subtracted each image from the convolved reference
frame. The fluxes of the AGN and control stars to estimate errors were
determined by fitting a PSF-weighted aperture over each source, thus
producing a differential light curve.

\subsubsection{Construction of Light Curves}
\label{section:lightcurves}

\begin{figure*}
\centering
\includegraphics[width=0.9\textwidth]{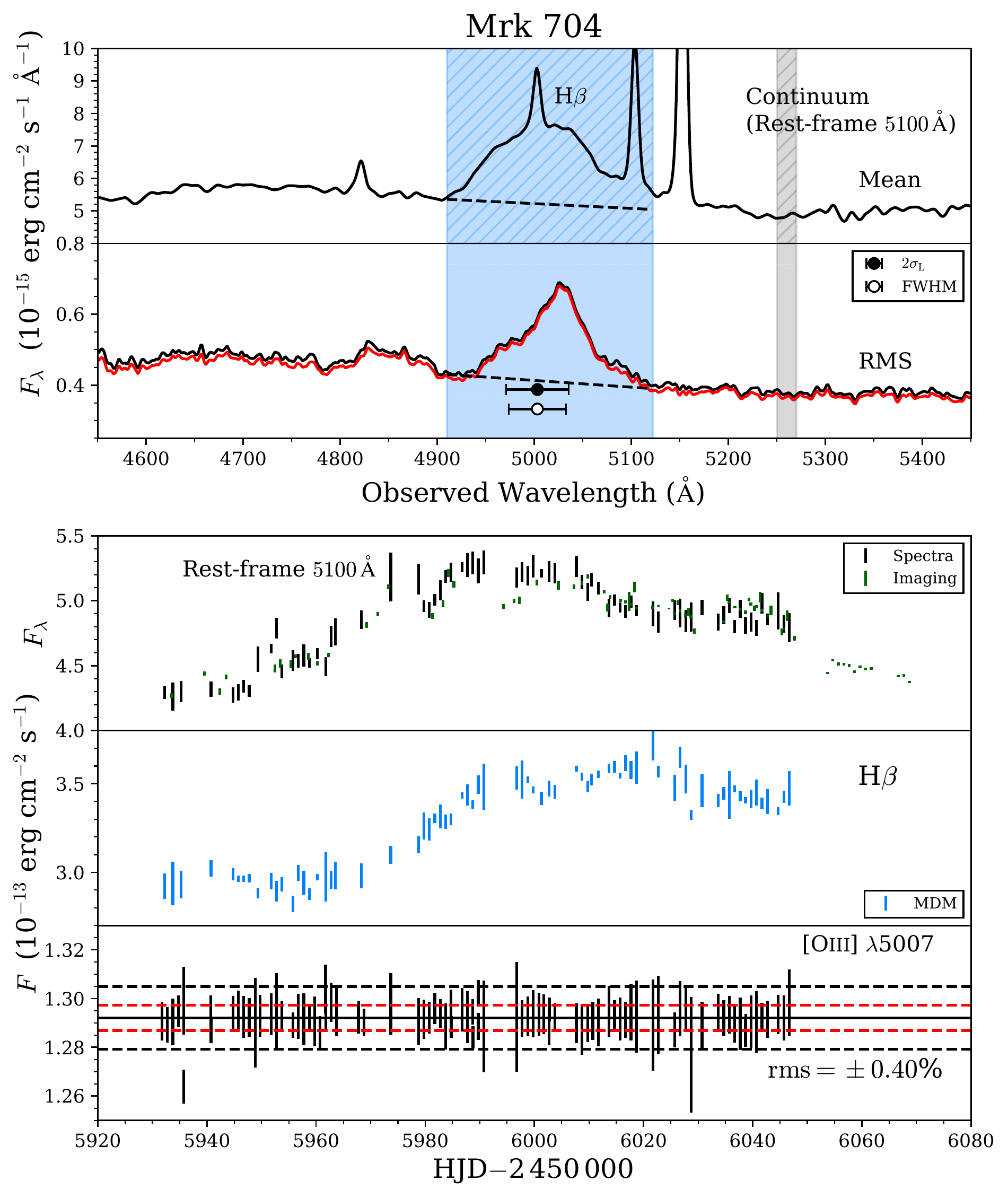}
\caption{The top panel shows the weighted mean spectrum $\langle
  F(\lambda) \rangle$ (equation \ref{eq:meanspec}) of Mrk 704 in the
  observed frame based on the MDM spectra.  The blue shaded region
  shows the integration range for H$\beta$ and the dashed line
  underneath shows the underlying continuum assumed in the line
  integration.  The 5100\,\AA\ continuum measurement is the average
  flux in the gray-shaded region.  The second panel shows the RMS
  spectrum $\sigma_{\rm rms}(\lambda)$ (equation \ref{eq:RMSspec}) in
  black, and the intrinsic variability $\sigma_{\rm var}(\lambda)$
  (equation \ref{eq:intrinsicRMSspec}) in red.  The errorbars show the
  rms linewidth ($\sigma_{\rm L}$) and full-width at half maximum
  (FWHM).  We note that \heii\,$\lambda4686$ also appears in the RMS
  residual spectrum; a more sophisticated analysis will be required to
  separate the \heii\ emission from blended \feii\ emission and
  features in the host-galaxy spectrum.  The lower three panels are,
  from top to bottom, the light curves for the 5100\,\AA\ continuum,
  the H$\beta$ emission line, and the [\oiii]\,$\lambda 5007$ narrow
  emission line, with the latter used as a measure of the fidelity of
  the flux calibration. In the bottom panel, red dashed lines indicate
  the $1\sigma$ scatter, while the black dashed lines indicate $\pm
  1\%$ of the mean flux.}
\label{fig:mrk704data}
\end{figure*}

\begin{figure*}
\centering
\includegraphics[width=0.9\textwidth]{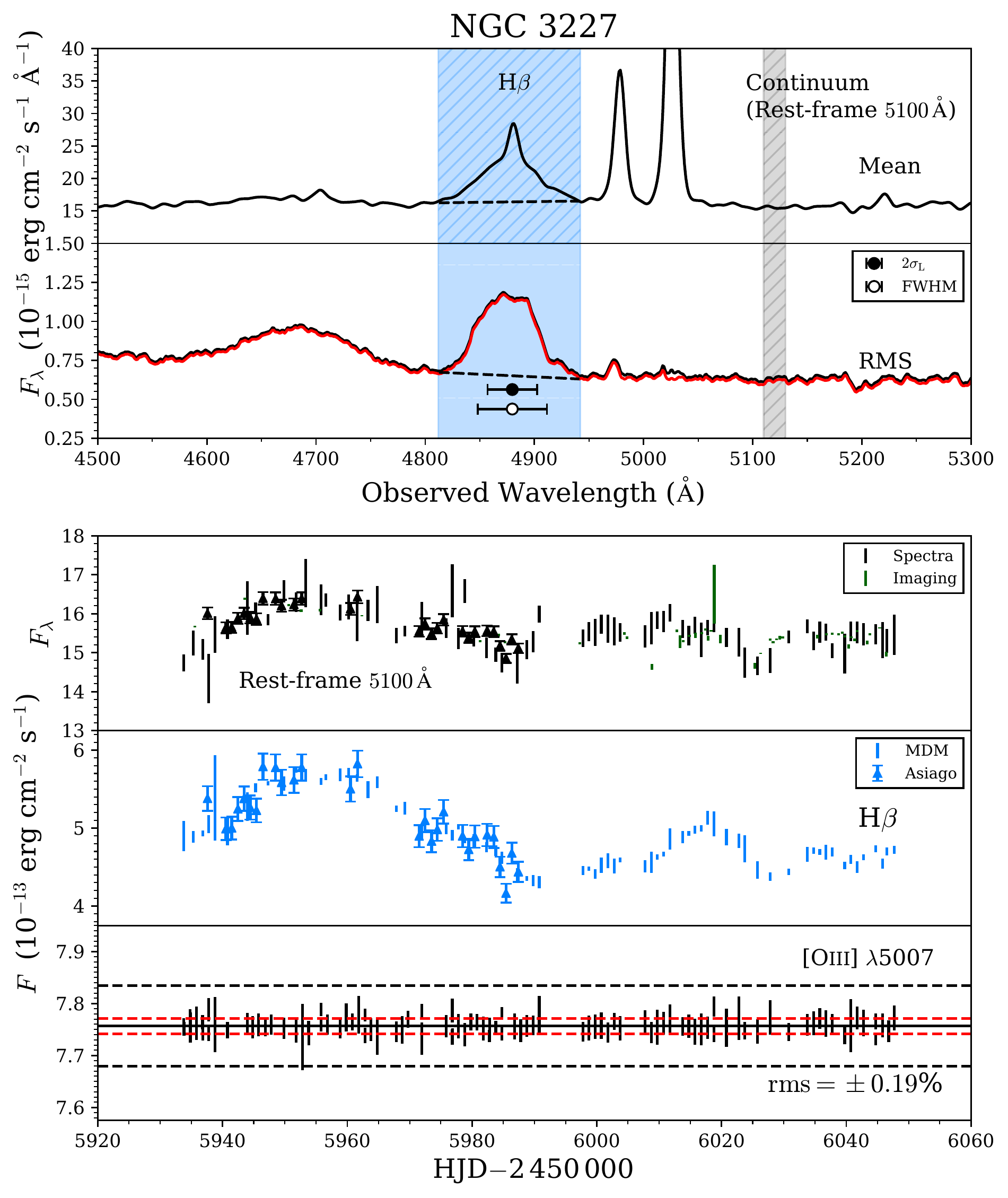}
\caption{Mean and RMS spectra for the 2012 observtions of NGC 3227 and
  the 5100\,\AA\ continuum, H$\beta$, and [\oiii]\,$\lambda5007$ light
  curves.  The format is the same as in Figure \ref{fig:mrk704data}.}
\label{fig:n3227data}
\end{figure*}

\begin{figure*}
\centering
\includegraphics[width=0.9\textwidth]{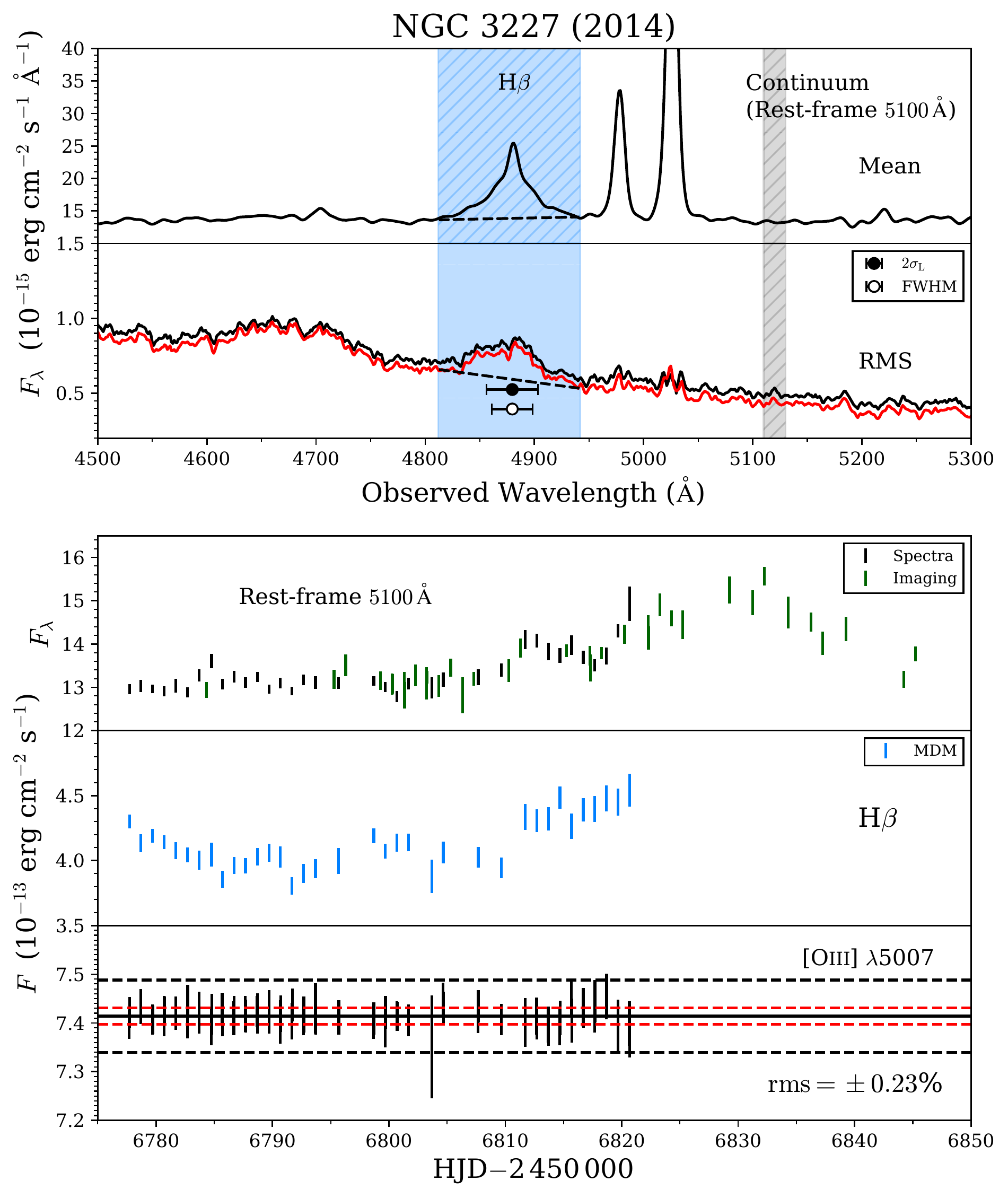}
\caption{Mean and RMS spectra for the 2014 observations of NGC 3227
  and the 5100\,\AA\ continuum, H$\beta$, and [\oiii]\,$\lambda5007$
  light curves.  The format is the same as in Figure
  \ref{fig:mrk704data}.}
\label{fig:n3227_2014data}
\end{figure*}

\begin{figure*}
\centering
\includegraphics[width=0.9\textwidth]{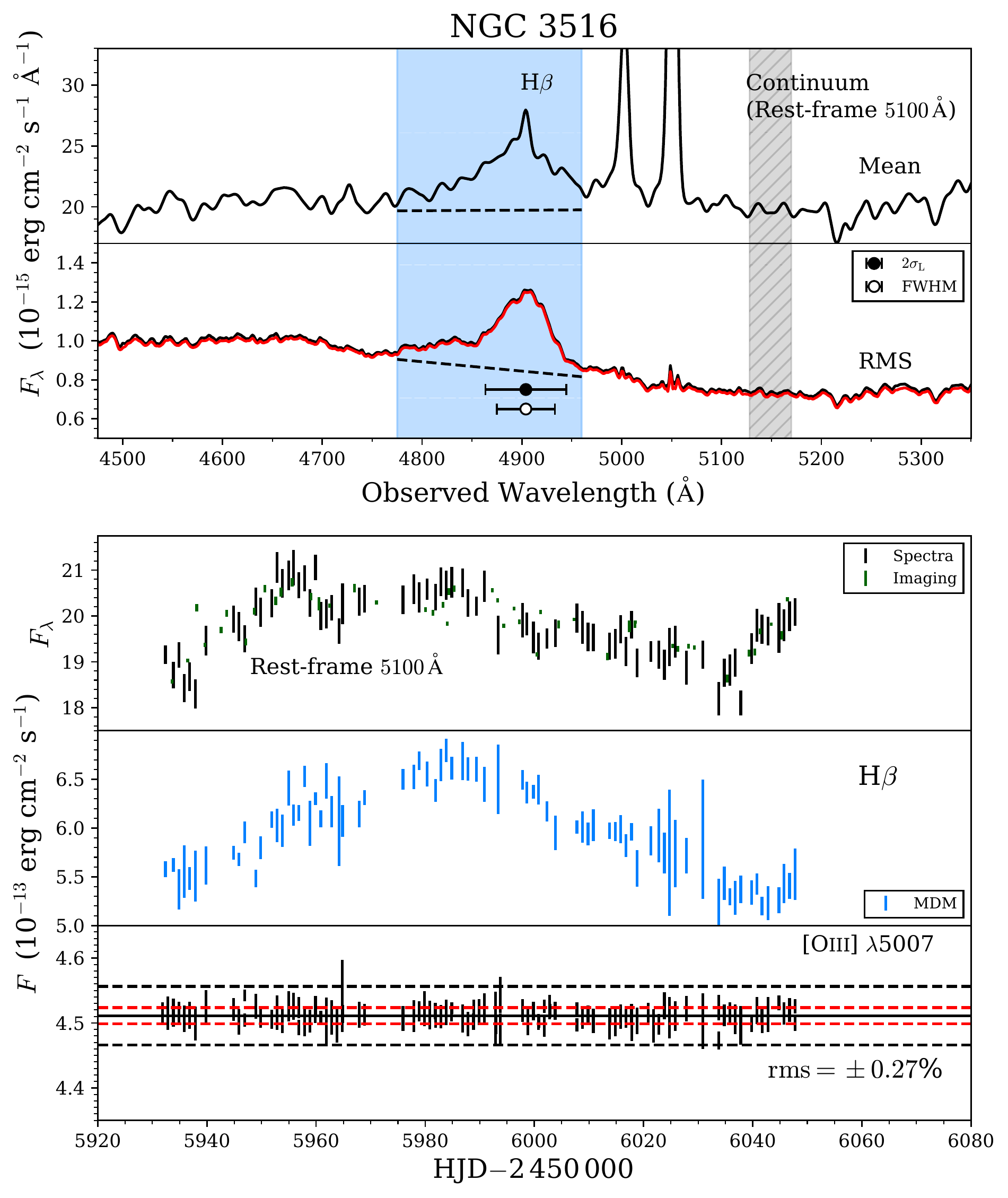}
\caption{Mean and RMS spectra for NGC 3516 and the 5100\,\AA\
  continuum, H$\beta$, and [\oiii]\,$\lambda5007$ light curves.  The
  format is the same as in Figure \ref{fig:mrk704data}.}
\label{fig:n3516data}
\end{figure*}

\begin{figure*}
\centering
\includegraphics[width=0.9\textwidth]{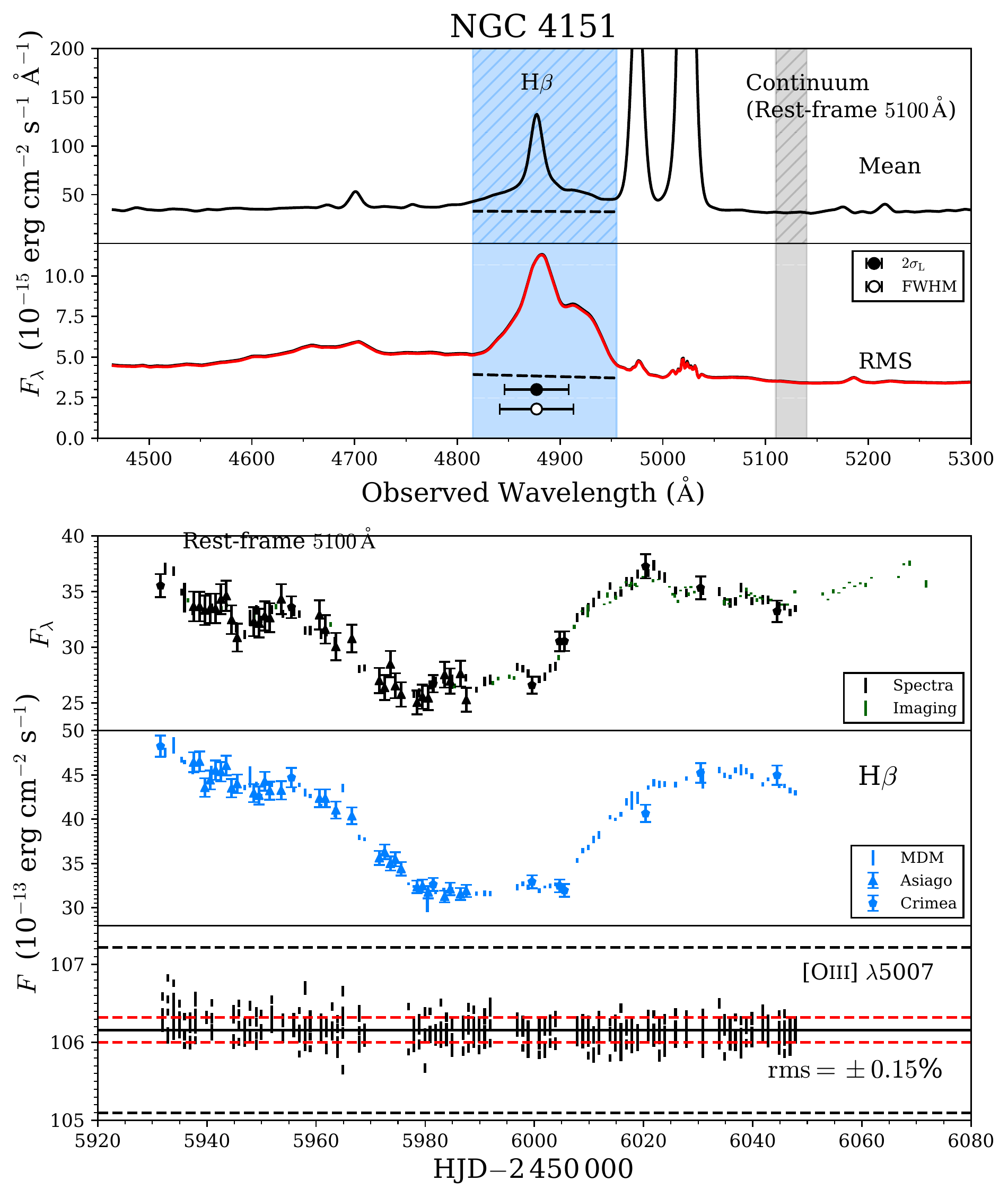}
\caption{Mean and RMS spectra for NGC 4151 and the 5100\,\AA\
  continuum, H$\beta$, and [\oiii]\,$\lambda5007$ light curves.  The
  format is the same as in Figure \ref{fig:mrk704data}.}
\label{fig:n4151data}
\end{figure*}

\begin{figure*}
\centering
\includegraphics[width=0.9\textwidth]{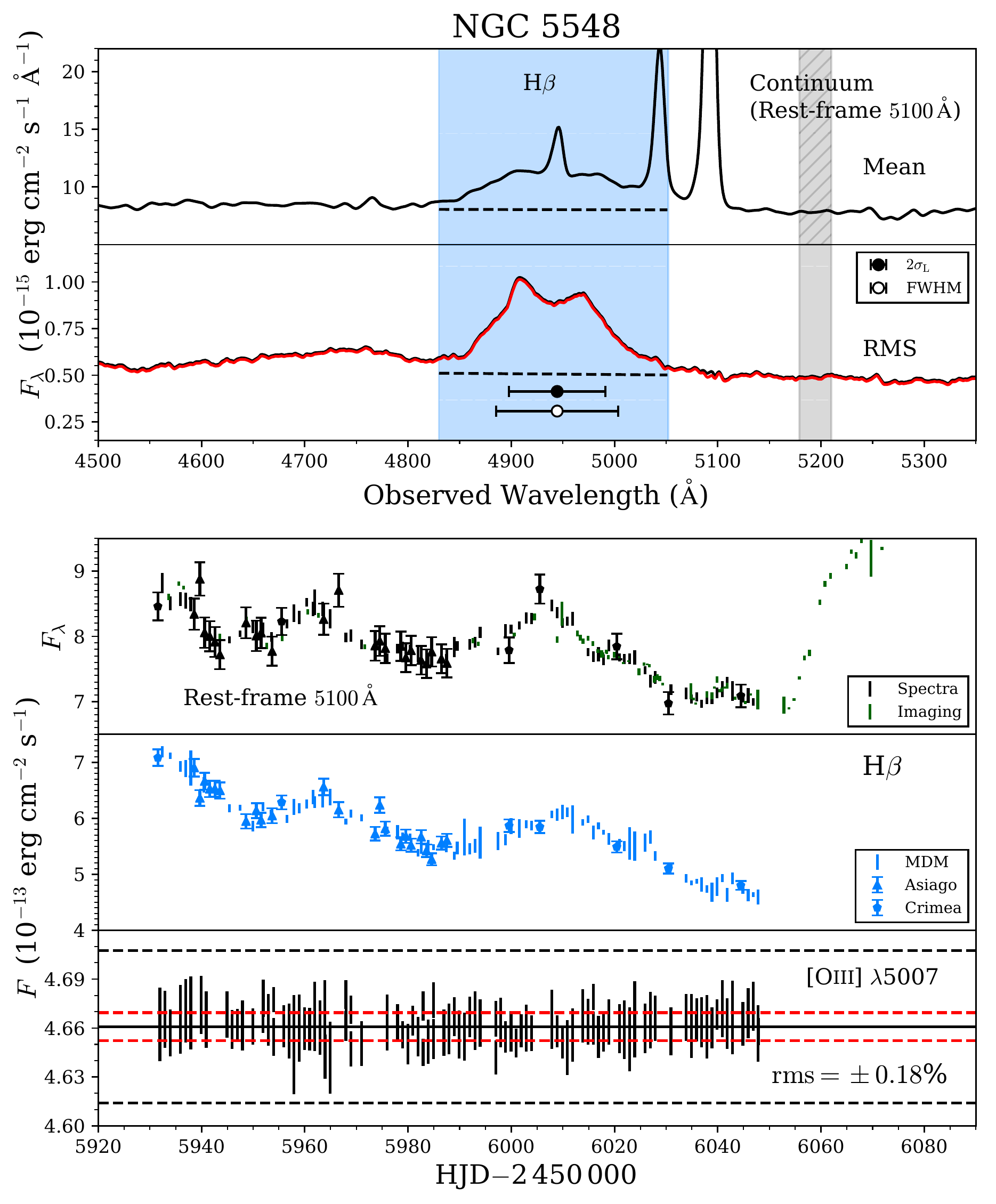}
\caption{Mean and RMS spectra for NGC 5548 and the 5100\,\AA\ continuum,
H$\beta$, and  [\oiii]\,$\lambda5007$ light curves.  The format is the same as in Figure \ref{fig:mrk704data}.}
\label{fig:n5548data}
\end{figure*}

A spectroscopic continuum light curve, nominally at $\sim5100$\,\AA\
in the rest-frame of each AGN, is formed by averaging the flux
densities over the wavelength ranges given in Table
3 
and shown as a shaded region in Figures \ref{fig:mrk704data} --
\ref{fig:n5548data}.  Our final continuum light curves are constructed
by merging the differential $V$-band light curves 
with the 5100\,\AA\ spectroscopic light curve by scaling multiplicatively (to match the
variations) and shifting additively (to account for the different mean
flux levels in each reference image) the differential continuum light
curves. We found that the uncertainties on the
differential light curves are systematically too small because {\tt
  ISIS} takes into account only Poisson errors. To bypass this problem,
we rescale the errors based on measurements of other stars in the field
of view, as described in detail by \cite{Fausnaugh16}.

The emission-line light curves are generated by interpolating a simple
linear continuum underneath the emission lines using the windows 
given in Table 4 and integrating the
flux above this continuum between the limits given in Table
3 
and illustrated in Figures \ref{fig:mrk704data} --
\ref{fig:n5548data}.  These measurements are fairly crude, but are
intended to capture the emission-line variations as opposed to all the
emission-line flux.  A more sophisticated treatment is deferred to a
future paper. We estimate the uncertainties using a local linear interpolation 
method described in detail by \cite{Fausnaugh17b}, which rescales the 
statistical uncertainties of the light curves so that they are consistent with 
the observed night to night scatter.

The final continuum and emission line light curves are given in Tables 5
and 6 respectively.
All the light curves are shown in Figures \ref{fig:mrk704data} --
\ref{fig:n5548data}.  The statistical properties of the light curves
are summarized in Table 7, including the number of observations
$N_{\rm obs}$, median cadence $\Delta t_{\rm med}$, mean flux $\langle
F \rangle$, the mean signal-to-noise ratio $\langle S/N \rangle$, the excess
variance $F_{\rm var} = \sigma_{\rm var}/\hat F$, where $\sigma_{\rm
  var}$ and $\hat F$ are determined in the same way as in Equation
\ref{eq:intrinsicRMSspec} (after integrating over $\lambda$ to produce
the light curves), and the significance
\begin{align} {\rm (S/N)_{var}} = \frac{\sigma_{\rm var}}{\overline
      \sigma \sqrt{ 2/N_{\rm obs}}}
\end{align}
at which variability is detected, where $\overline \sigma$ is the mean measurement uncertainty.  Further
details can be found in \citet{Fausnaugh17b}.

\section {TIME-SERIES ANALYSIS}
\label{section:timeseries}

\subsection{Mean Emission-Line Lags}

Our initial goal is to determine the mean time scale for the response of
the H$\beta$ emission line to continuum variations, which we later
use to determine the mass of the central black hole.

The time series analysis is carried out using two common methodologies,
interpolated cross-correlation \citep{Gaskell86,Gaskell87,White94,Peterson98,
Peterson04} and the stochastic process modeling algorithm {\tt JAVELIN}\footnote{\url{http://www.astronomy.ohio-state.edu/~yingzu/codes.html\#javelin}}
 \citep{Zu11}. A more complete
description of how we have employed these methods for such analysis is
provided by \cite{Fausnaugh17b}.

Results of the time-series analysis are given in Table
8 
and shown graphically in the right-hand panels of Figures
\ref{fig:mrk704lagccf} -- \ref{fig:n5548lagccf}.  It is interesting to notice 
that the three AGNs from \cite{Denney09a}, re-observed in
this program, all have shorter lags than they did in 2007.  In the case
of NGC 3516, the factor-of-two decrease in the H$\beta$ lag is
consistent with the factor of four decrease in the AGN luminosity and
the expected scaling relation $R_{\rm BLR} \propto L_{\rm
  AGN}^{1/2}$. In the case of NGC 3227, the H$\beta$ lag also
decreased by a factor of two from 2007, but the AGN luminosity is in
fact slightly higher in 2012 and 2014.  In 2012, NGC 5548 had been in
a prolonged faint state for a few years \citep{Peterson13} and by
2013--14 heavy internal absorption became an important factor
\citep{Kaastra14,DeRosa15}. In both 2012 and 2014 \citep{Pei17}, the
H$\beta$ lag is found to be surprisingly short given the AGN
luminosity at the time. The implications of this are not yet clear,
although it appears that increased absorption within the BLR plays
some role.
The 2014 data on NGC 3227 are quite marginal,
and {\tt JAVELIN} was unable to converge on a solution for the lag. The ICCF analysis,
however, shows consistency with the 2012 results.
In the case of NGC 4151, the H$\beta$ lag is in good agreement with that obtained by \cite{Bentz06a}.

We note in passing that we also attempted to measure the variations of the
\heii\,$\lambda 4686$ line, which is clearly seen in the RMS residual spectra
of each source (Figs.\ \ref{fig:mrk704data} -- \ref{fig:n5548data}). Unfortunately
this is a weak, low contrast feature, and the measurements are
very uncertain on account of the difficulties in defining the underlying continuum.
Contamination of the spectra by the host-galaxy starlight is a significant
problem in low-luminosity AGNs, and it needs to be modeled and subtracted
for a reliable \heii\ measurement. We defer this to a future paper.

\subsection{Velocity-Resolved Lags}

\begin{figure*}
\centering
\includegraphics[width=0.9\textwidth]{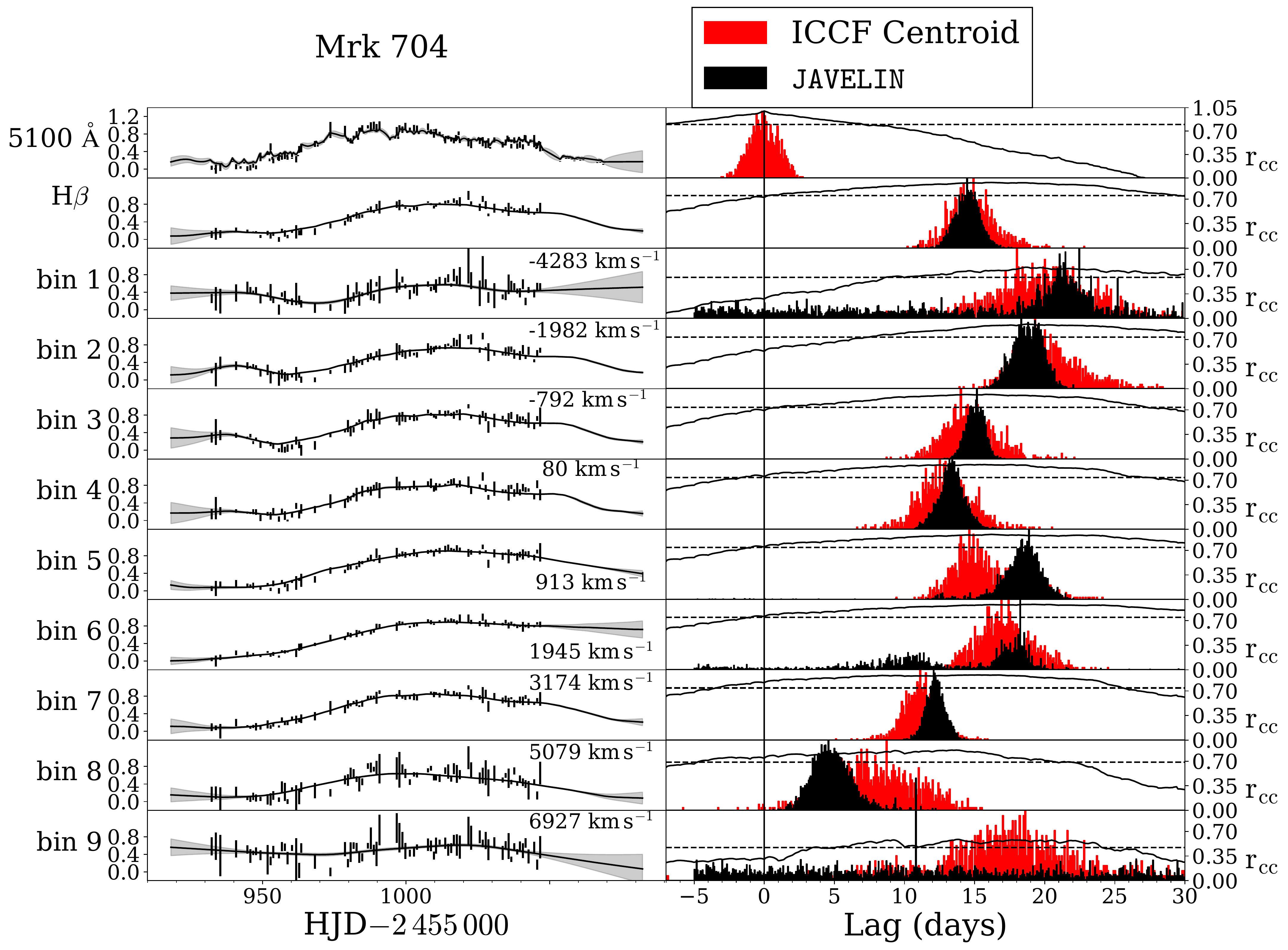}
\caption{Left-hand panels: Light curves for Mrk 704. The top panel shows
the 5100\,\AA\ continuum light curve and the integrated H$\beta$ light
curve is shown underneath. Underneath are light curves in different
Doppler velocity bins, starting with the far blueward wing and proceeding down
the the far redward wing, with the flux-weighted average velocity of the bin
labeled. Solid lines and shaded regions give the {\tt JAVELIN} models 
and the $1\sigma$ uncertainties.
Right-hand panels: Cross-correlations for Mrk 704.
The solid line shows the cross-correlation function generated by
cross-correlating the light curve to the immediate left with the 5100\,\AA\
continuum light curve in the upper left panel; the upper right panel is the
continuum autocorrelation function. The dashed lines are drawn at
$0.8r_{\rm max}$, where $r_{\rm max}$ is the peak of the cross-correlation
function, which occurs at $\tau_{\rm peak}$; values above this threshold are used to compute 
the centroid $\tau_{\rm cent}$. The cross-correlation centroid
distribution \citep[see][]{Peterson98} is shown in red and the 
 {\tt JAVELIN} posterior distribution of lags is shown in black.}
\label{fig:mrk704lagccf}
\end{figure*}

\begin{figure*}
\centering
\includegraphics[width=0.9\textwidth]{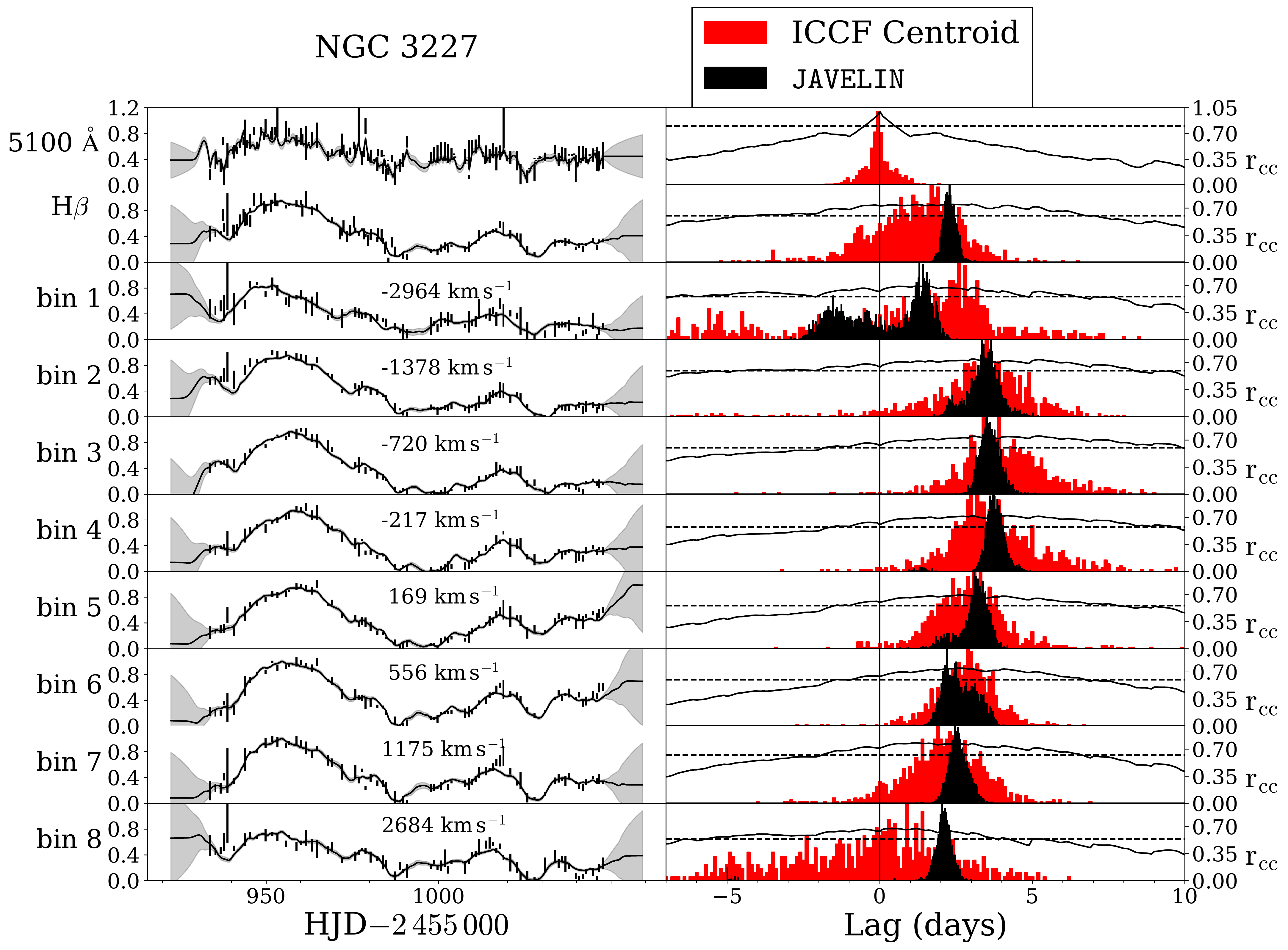}
\caption{Light curves and cross-correlation functions for NGC 3227.
  The format is the same as in Figure \ref{fig:mrk704lagccf}.}
\label{fig:n3227lagccf}
\end{figure*}

\begin{figure*}
\centering
\includegraphics[width=0.9\textwidth]{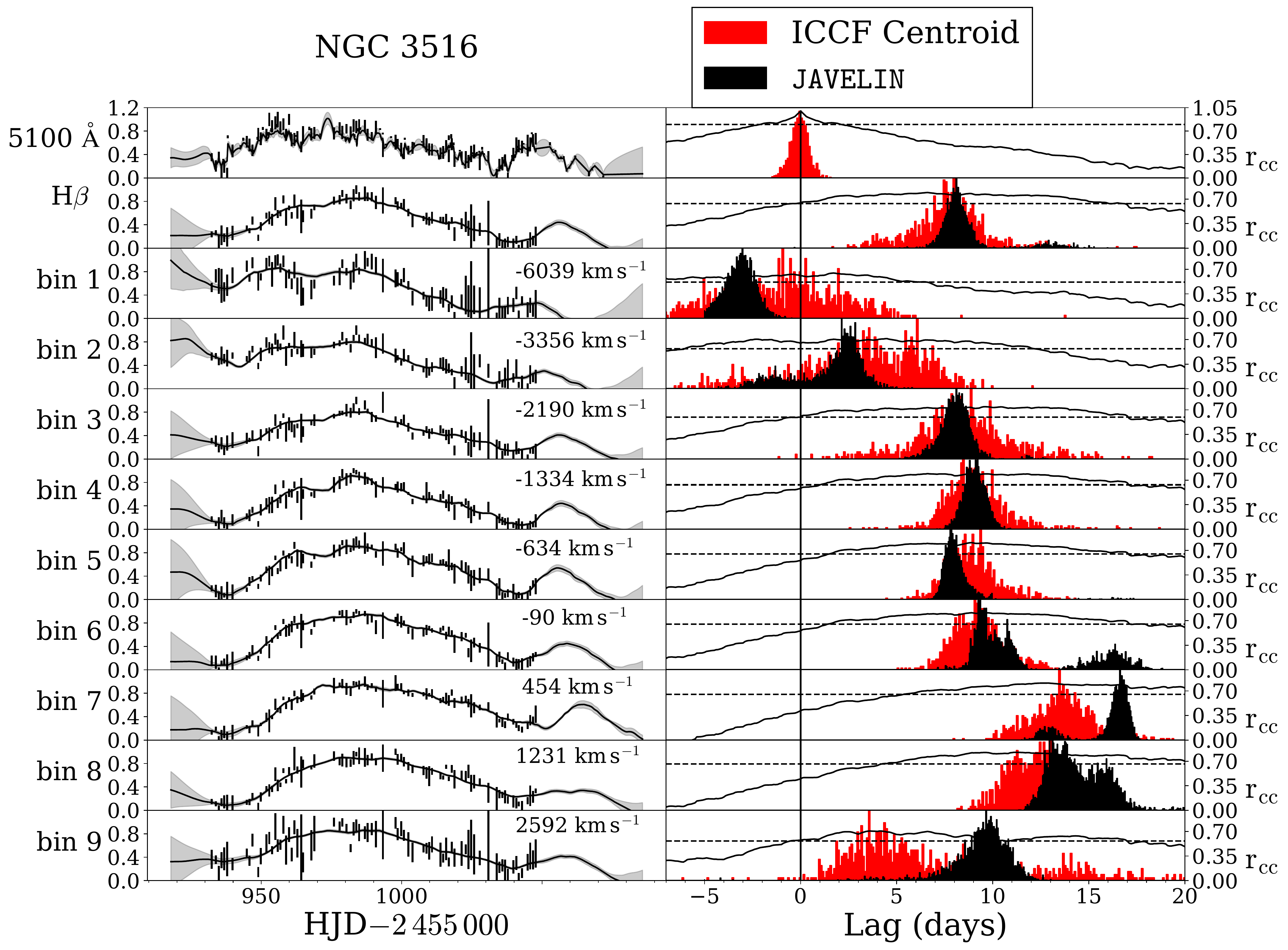}
\caption{Light curves and cross-correlation functions for NGC 3516.
  The format is the same as in Figure \ref{fig:mrk704lagccf}.}
\label{fig:n3516lagccf}
\end{figure*}

\begin{figure*}
\centering
\includegraphics[width=0.9\textwidth]{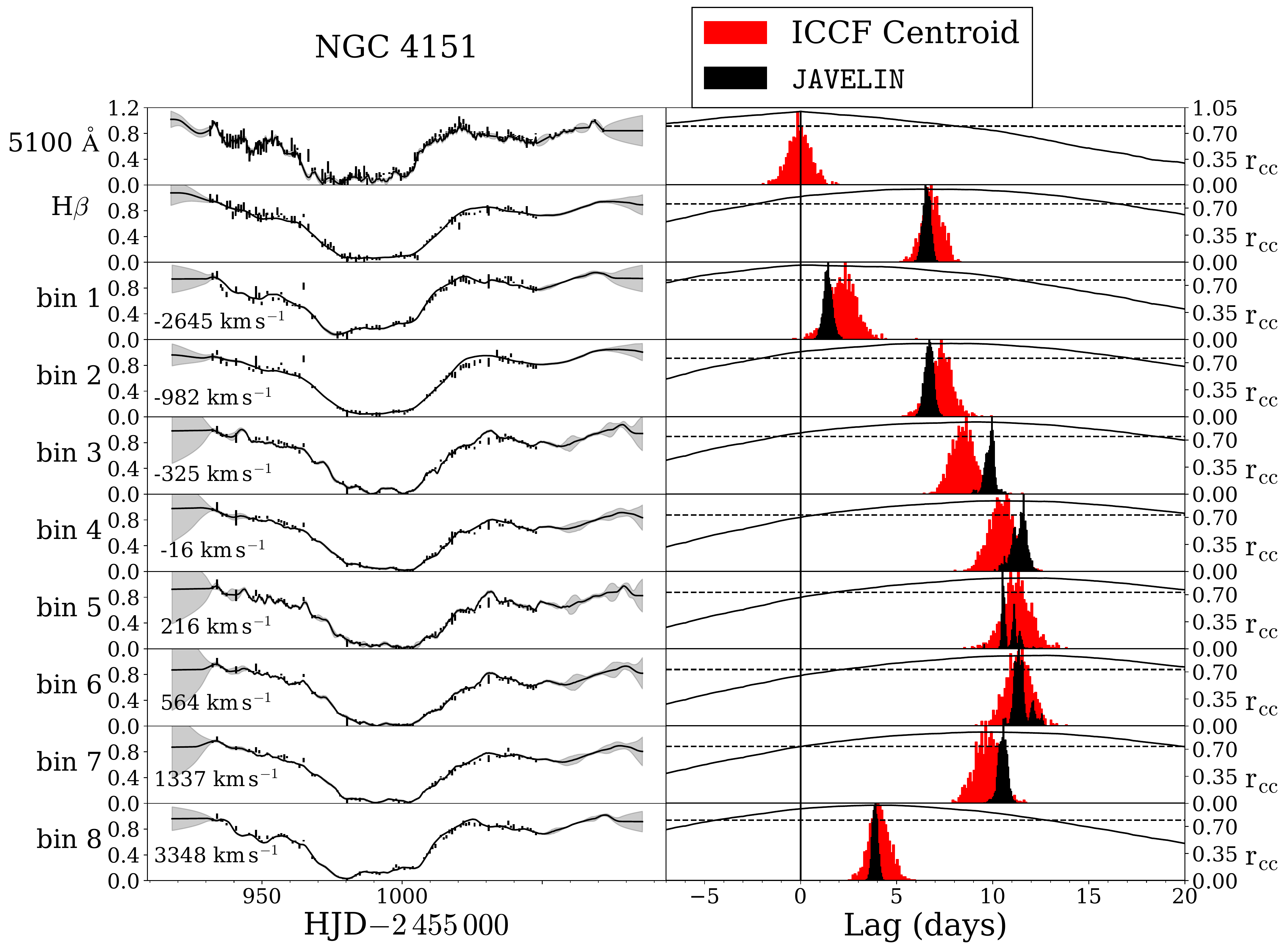}
\caption{Light curves and cross-correlation functions for NGC 4151.
  The format is the same as in Figure \ref{fig:mrk704lagccf}.}
\label{fig:n4151lagccf}
\end{figure*}

\begin{figure*}
\centering
\includegraphics[width=0.9\textwidth]{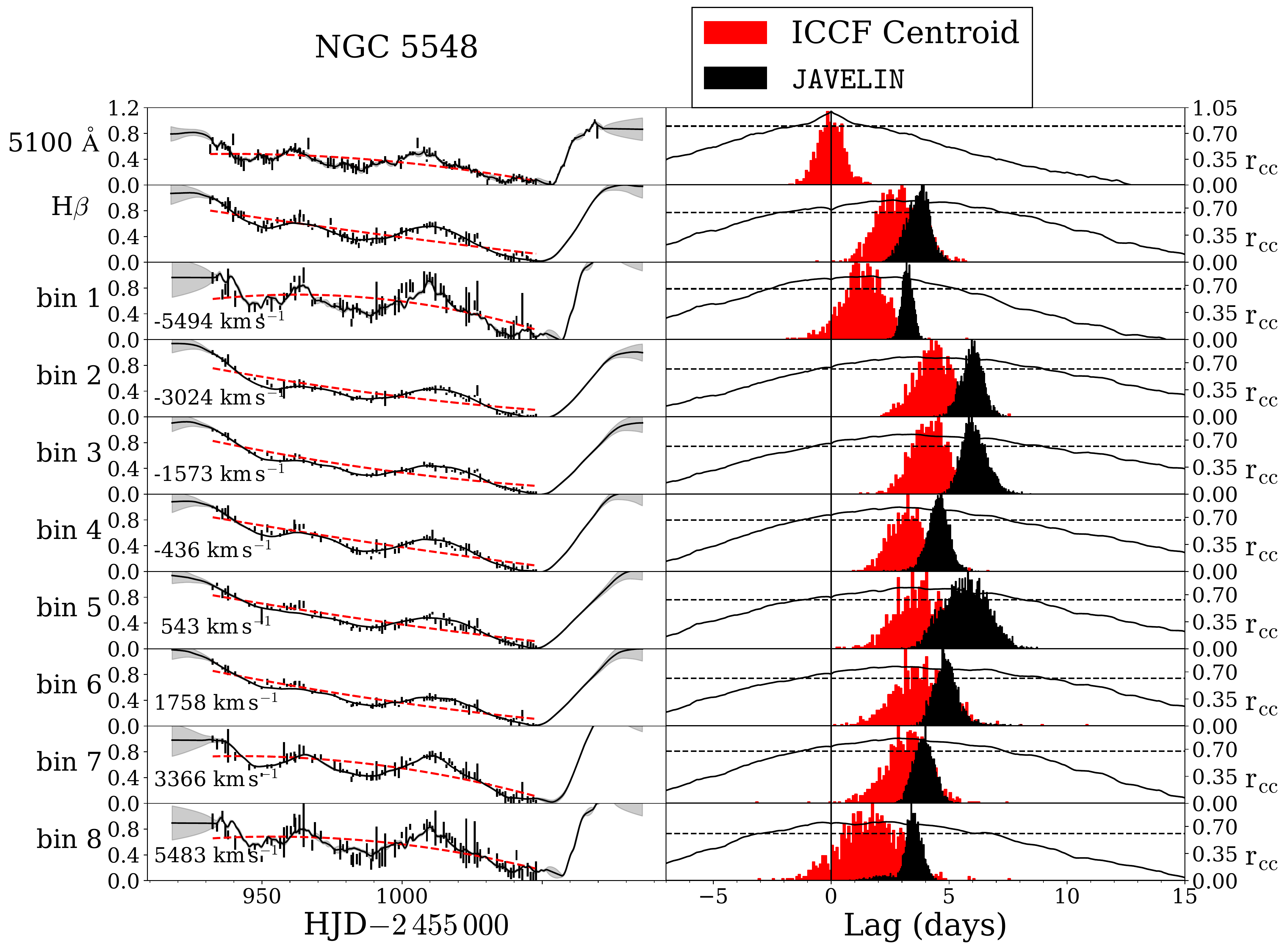}
\caption{Light curves and cross-correlation functions for NGC 5548.
  The format is the same as in Figure \ref{fig:mrk704lagccf}.  The
  dashed red lines show the results of a second-order polynomial
  linear least-squares fits, which were used to detrend the light
  curves prior to calculating the ICCF.}
\label{fig:n5548lagccf}
\end{figure*}

\begin{figure*}
\centering
\includegraphics[width=0.9\textwidth]{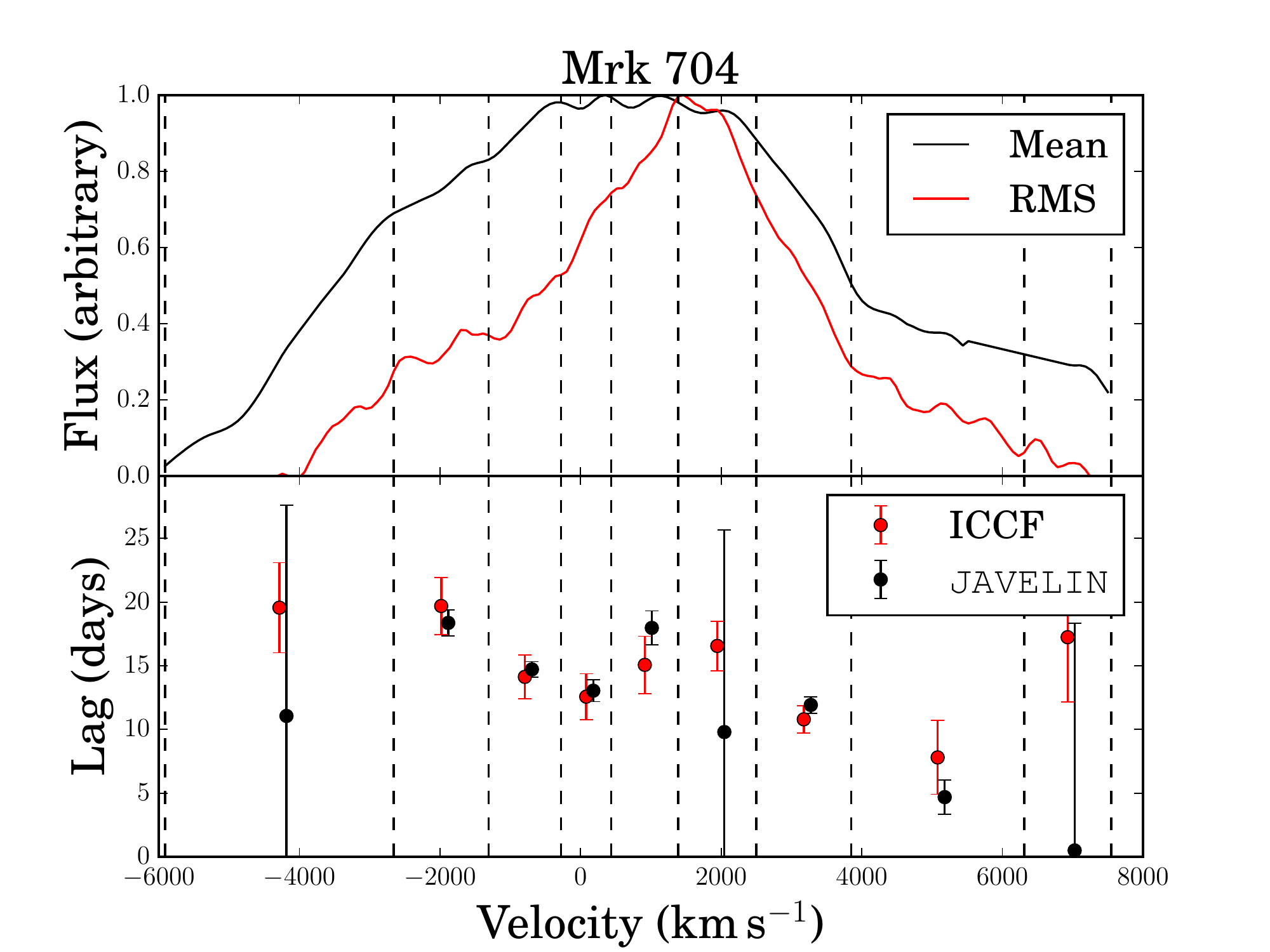}
\caption{Top panel: Mean (black) and RMS residual (red) H$\beta$
  profiles for Mrk 704. The narrow-line component of H$\beta$ and the
  [O\,{\sc iii}]\,$\lambda\lambda4959$, 5007 lines have been modeled
  out of the mean spectrum.
  The vertical dashed lines show the velocity
  bins used to produce velocity-resolved light curves. 
  The bin boundaries were chosen so that the
  total mean-spectrum flux in each bin is approximately the
  same. Bottom panel: Lags measured for the emission in each velocity
  bin, with interpolated cross-correlation function (ICCF) lags shown
  in red and {\tt JAVELIN} lags shown in black.  The {\tt JAVELIN}
  lags are offset by +100 km\,s$^{-1}$ for clarity.}
\label{fig:mrk704velres}
\end{figure*}

\begin{figure*}
\centering
\includegraphics[width=0.9\textwidth]{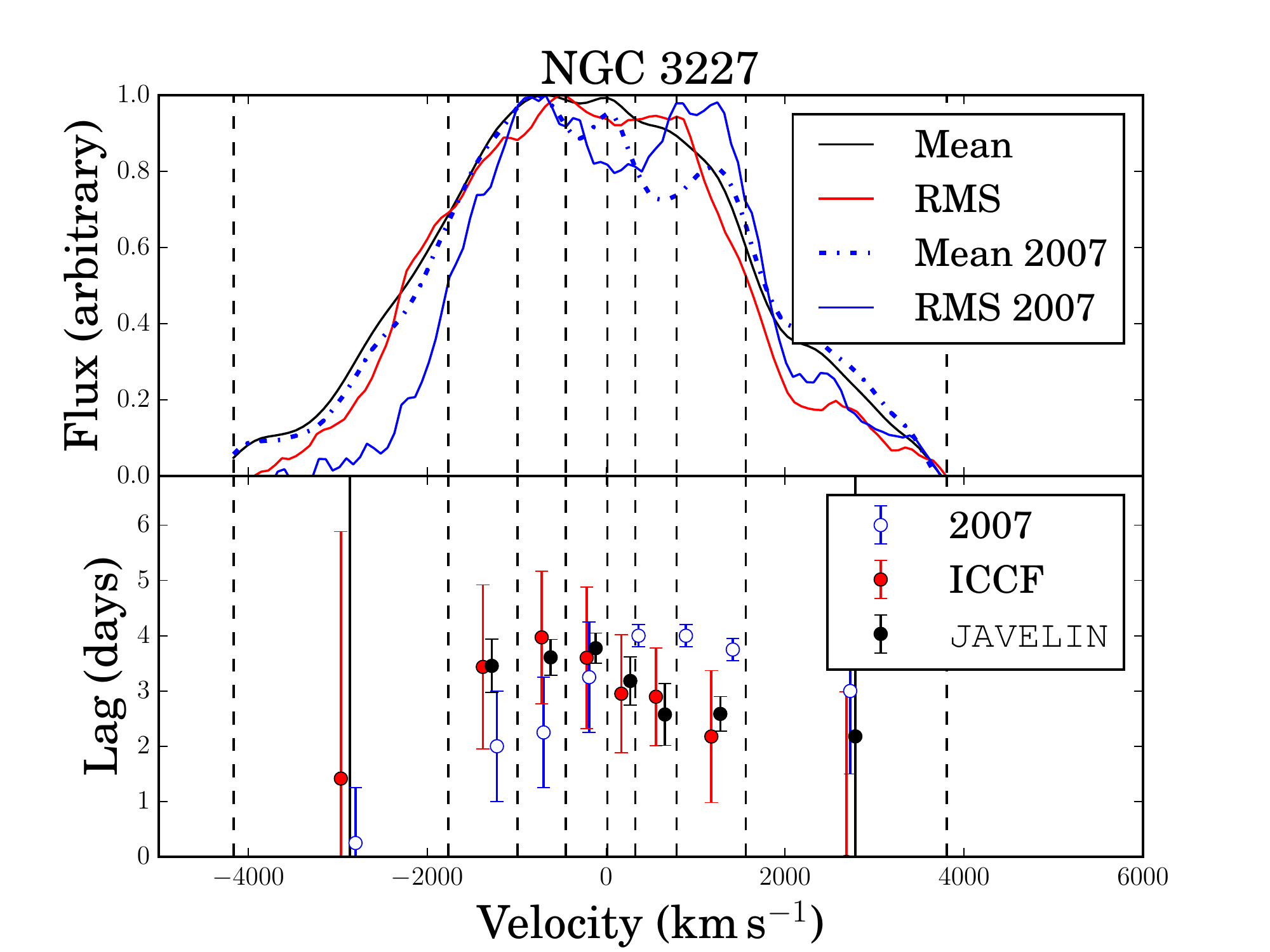}
\caption{Upper panel shows the mean and RMS residuals for H$\beta$ in
  NGC 3227 and lower panel shows the lag in each velocity bin.  The
  format is the same as in Fig. \ref{fig:mrk704velres}. 
  In the upper panel, the 2007 mean (blue dashed line) and rms (blue solid line)
  H$\beta$ profiles from \cite{Denney09b} are shown. In the lower panel
  the lags from \cite{Denney09a} are shown 
  as open blue circles. Note that \cite{Denney09a} used slightly different 
  velocity bins than those defined here.}
\label{fig:n3227velres}
\end{figure*}

\begin{figure*}
\centering
\includegraphics[width=0.9\textwidth]{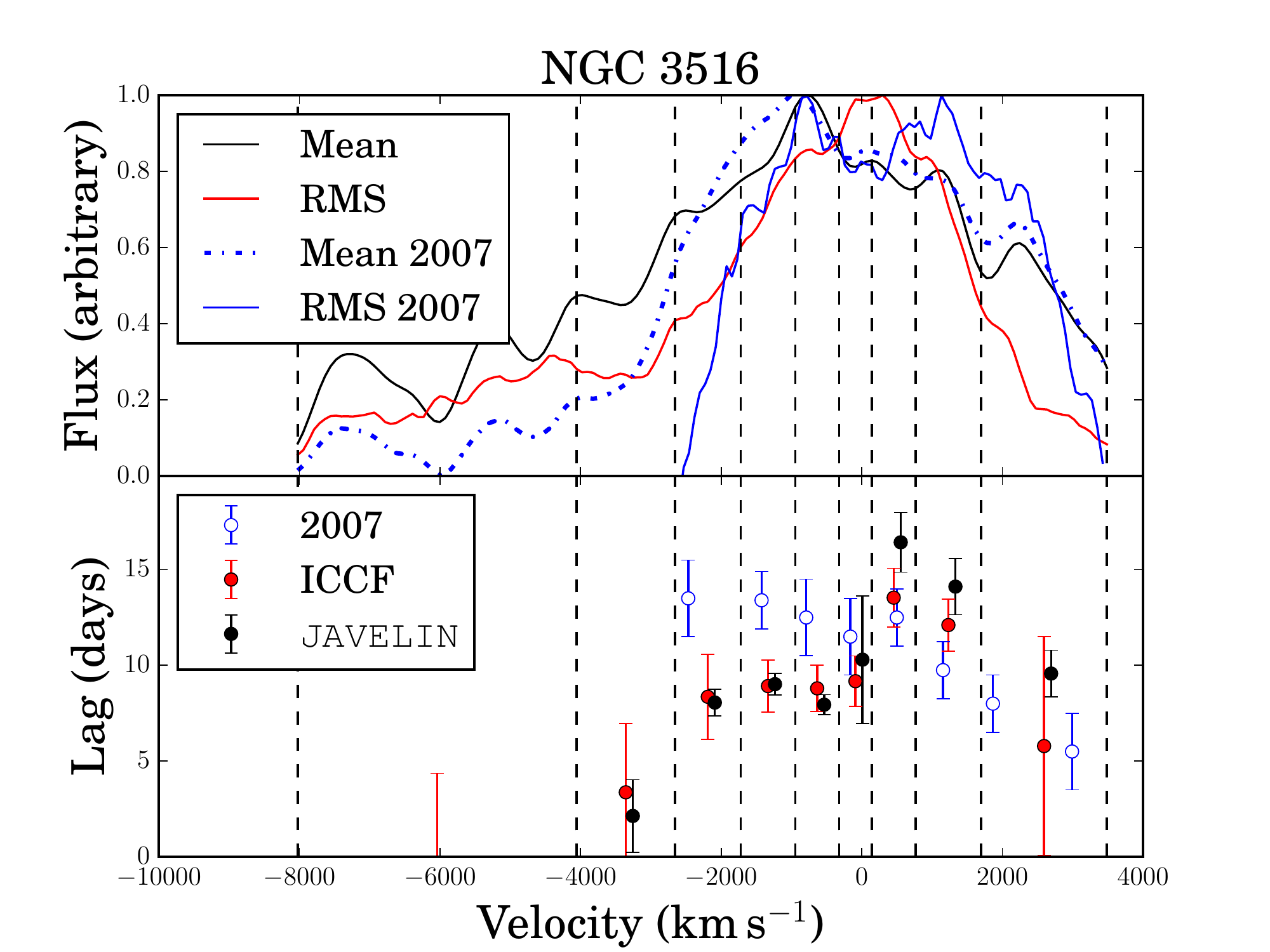}
\caption{Upper panel shows the mean and RMS residuals for H$\beta$ in
  NGC 3516 and lower panel shows the lag in each velocity bin.  The
  format is the same as in Fig. \ref{fig:mrk704velres}. 
 In the upper panel, the 2007 mean (blue dashed line) and rms (blue solid line)
  H$\beta$ profiles from \cite{Denney09b} are shown. In the lower panel
  the lags from \cite{Denney09a} are shown 
  as open blue circles.}
\label{fig:n3516velres}
\end{figure*}

\begin{figure*}
\centering
\includegraphics[width=0.9\textwidth]{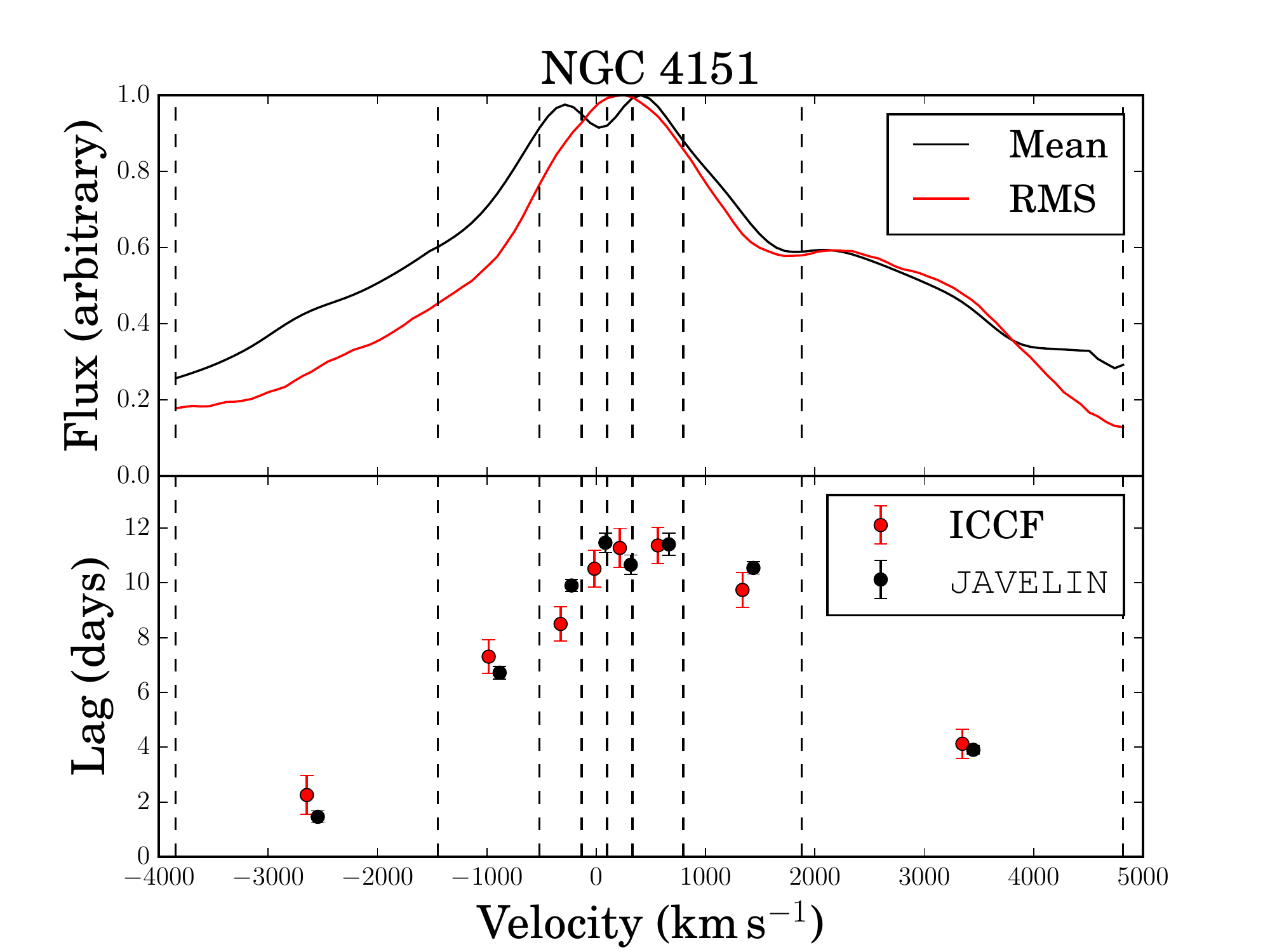}
\caption{Upper panel shows the mean and RMS residuals for H$\beta$ in
  NGC 4151 and lower panel shows the lag in each velocity bin.  The
  format is the same as in Fig. \ref{fig:mrk704velres}.}
\label{fig:n4151velres}
\end{figure*}

\begin{figure*}
\centering
\includegraphics[width=0.9\textwidth]{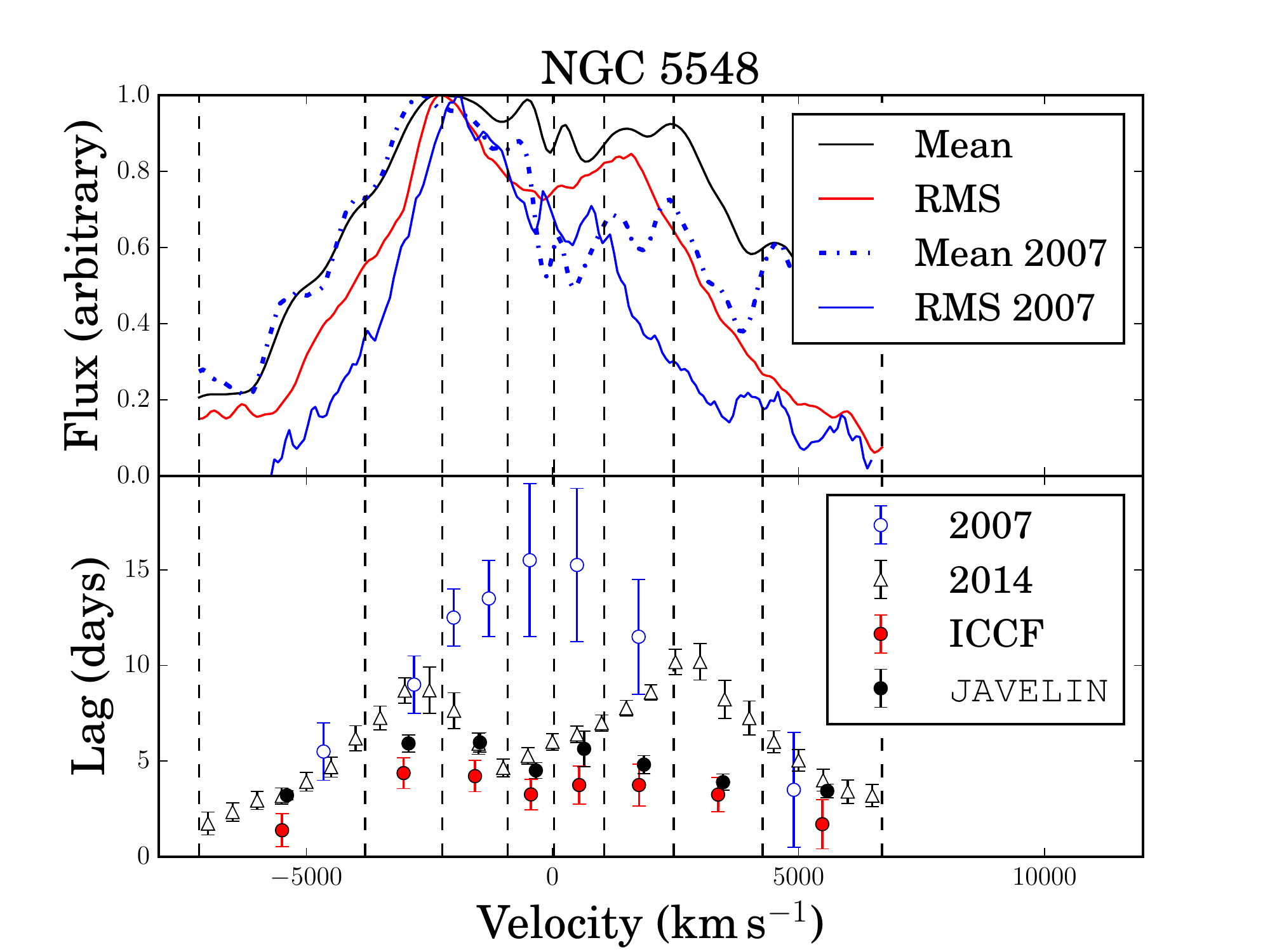}
\caption{Upper panel shows the mean and RMS residuals for H$\beta$ in
  NGC 5548 and lower panel shows the lag in each velocity bin.  The
  format is the same as in Fig. \ref{fig:mrk704velres}. 
  In the upper panel, the 2007 mean (blue dashed line) and rms (blue solid line)
  H$\beta$ profiles from \cite{Denney09b} are shown. In the lower panel
  the lags from \cite{Denney09a} are shown 
  as open blue circles,
  while the ones from \cite{Pei17} are shown was open black triangles. }
\label{fig:n5548velres}
\end{figure*}

The individual spectra are of high enough $S/N$ ratio and sufficiently
well-sampled in time that we can also divide each emission line into
line-of-sight velocity bins to see if there are any indications of
gross kinematic signature and, in the cases of NGC 3227, NGC 3516, and
NGC 5548, compare these results with those obtained by \cite{Denney09a}. 
This is not a foolproof method of discerning the velocity field of the BLR
as experience has shown that reverberation effects are quite subtle
and attempting to characterize an individual velocity bin with a
single average lag could be misleading. While we must interpret the
results cautiously, detection of a velocity-dependent lag signature
identifies good candidates for more ambitious attempts to determine
the BLR structure and velocity field by either forward modeling
\citep{Pancoast12,Pancoast14, Grier17} or regularization
\citep{Horne04,Bentz10b,Grier13a,Skielboe2015}.
The results of measuring velocity-dependent lags are shown
in the lower panels of Figs. \ref{fig:mrk704lagccf} -- \ref{fig:n5548lagccf}
 in a format that can be easily compared with
Fig.\ 3 of \cite{Denney09a} in Figs. \ref{fig:mrk704velres} -- \ref{fig:n5548velres}.
We comment on each source individually:

\paragraph{Mrk 704}  Fig.\ \ref{fig:mrk704lagccf} shows that the highest
velocity blue-shifted and red-shifted bins have large lag uncertainties, so we
will disregard these. The remaining bins (Fig.\ \ref{fig:mrk704velres}) show a local lag minimum 
around line center ($V = 0$). At higher red-shifted velocities, the lags increase
to a maximum at $\sim 2000$\,\kms, then become smaller in the far wings.
On the blue-shifted side, the lags also increase, but we do not see a turnover
toward smaller lags at higher velocity.
A similar pattern with relatively small lags at line center compared to the
wings, is seen in NGC 5548 in 2014 in H$\beta$ \citep{Pei17}
and also probably in Ly$\alpha$ \citep{DeRosa15}. The BLR velocity
field in Mrk 704 may well have multiple components, and requires more 
sophisticated modeling.

\paragraph{NGC 3227} As with Mrk 704, the highest velocity bins
have large errors and should be disregarded (Fig.\ \ref{fig:n3227lagccf}). The
remaining bins show a pattern that suggests a virialized BLR
(Fig.\ \ref{fig:n3227velres}), with large lags at line center and shorter
lags at higher positive and negative velocities. This can be compared with
results from \cite{Denney09a}, which do not show a decrease
in the lag at higher positive velocities.  Lower lags at high negative
velocity might be interpreted as evidence for outflow. There is no strong evidence for 
outflow in the 2012 data. Again, more sophisticated modeling will
clarify the situation. As noted earlier, the 2014 data on this source are
 marginal and are not included in this analysis. We note that a very similar
 dependence of lag on velocity bin is seen from an independent RM campaign from
 2017 (M.S.\ Brotherton, private communication).

\paragraph{NGC 3516} In 2007 \citep{Denney09a}, the highest positive velocities in the
H$\beta$ emission line showed the shortest lags, with the lags steadily
increasing toward line center and continuing to increase slightly to
higher negative velocity  (Fig.\ \ref{fig:n3516velres}). 
This behavior could be interpreted as an infall
signature.  In 2012, at least in the core of the line, this trend
seems to be reversed.

\paragraph{NGC 4151} On account of the brightness and favorable
variability characteristics of NGC 4151 during this campaign, the results for this AGN are superb.
The uncertainties in the lag for each velocity bin are quite small
(Fig.\ \ref{fig:n4151lagccf}) and there is a very clear virial-like
pattern where the largest lags are seen at the lowest velocities  (Fig.\ \ref{fig:n4151velres}).

\paragraph{NGC 5548} Due to less favorable variability characteristics
in 2012, the NGC 5548 results are not as clear as they were in either
2007 \citep{Denney09a} or in 2014 \citep{Pei17}; the uncertainties in each velocity bin are
comparatively large (Fig. \ref{fig:n5548lagccf}).  The pattern as a
function of wavelength seems to be quite similar to the complex
pattern observed in 2014 (see Fig.\ 10 of \citealt{Pei17}) as well as
in 2015 \citealt{Lu16}. This is also similar to Mrk 704 (Fig.\ \ref{fig:mrk704velres}), 
and possibly indicates a multicomponent BLR.

\section {LINE WIDTH AND BLACK HOLE MASS CALCULATION}
\label{section:linewidths}

In order to compute the mass of the central black hole from Eq.\ (\ref{eq:mass}),
we need to characterize the line width $\Delta V$ in addition to the
mean emission-line lag $\tau$. The two line-width measures
commonly used for this are full width at half maximum (FWHM) and
the line dispersion
\begin{equation}
\sigma_{\rm line} = \left[ \frac{ \int v^2 P(v) dv}{\int P(v)\,dv}\right]^{1/2},
\end{equation}
 which is the square root of the second moment of the line. 
 The integral is over the line profile $P(v)$ as a function of line-of-sight (Doppler)
velocity. There are practical advantages and disadvantages to 
each of these.  The FWHM is usually trivial to measure, but presents problems when
the data are noisy or the profiles are complex. The line dispersion, on the other hand, requires attention
to blending with other features, but is computationally well-defined for any profile. 
The two measures
are not interchangeable, as their ratio varies with line shape, which is correlated with line width. There are compelling, but
not conclusive, arguments that line dispersion is the better parameter for
computing masses \citep{Denney13,Peterson14}, so we use $\sigma_{\rm line}$  in our mass
calculations, but report both measures for both the mean and RMS spectra in 
Table 9. 

For the mass calculation, we use $\sigma_{\rm line}$ from RMS spectra
as the line-width measure because the RMS profile reflects the motions of
the gas that is actually responding to the continuum flux variations. 
For the time delay, we use $\tau_{\tt JAV}$, though the uncertainties in
this quantity depend strongly on the assumption that all errors are Gaussian.
In Table 10, we list the time lags and line widths adopted for each data set,
and combine these to form the virial product VP. To put the virial products
on a calibrated mass scale, we 
adopt a mean scale factor of $f = 4.47 \pm 1.25$ \citep{Woo15}. The uncertainty in the scale factor
is propagated into the masses given in Table 10.

The virial product, ${\rm VP} = c\tau \Delta V^2/G$, is useful for
comparing the masses derived in different reverberation programs
because it involves only the two observables and physical
constants. While the VPs obtained here are in reasonable agreement with
earlier measurements (Columns 4 and 5 of Table 10), it is also clear
that the formal uncertainties derived from the
time delay and line width are too small.  There is clearly some
additional intrinsic scatter in the VP values beyond these formal
estimates, indicating
additional systematic uncertainties (perhaps due to the choice of
integration windows or blended spectral components) and/or
underestimated measure uncertainties.  The previous value
for NGC 5548 in the last row of Table 10 underscores this point: for
this entry, we used the mean and standard deviation from 16 previous
measurements of the VP based on H$\beta$ reverberation, spanning the
range $ 6.74 < \log {\rm VP} < 7.38$. The standard deviation of this
distribution is $\Delta \log {\rm VP} \approx 0.15$, which is probably
a good indicator of the true uncertainties in typical measurements. If
this is true generally, then the VP values measured here are all in
agreement with previous determinations.

\section{Conclusions}

We have presented new reverberation results for five bright local
Seyfert galaxies.  All five have been targets in previous
reverberation campaigns. In two cases,
Mrk~704 and NGC~4151, our previous campaigns
did not provide good measurements of the
emission-line lags or black hole masses.
Mrk~704 did not vary
in a fashion conducive to reverberation \citep{Barth15}, showing only
monotonically decreasing light curves. Our new data on NGC~4151 are far 
more extensive than those from our 2005 campaign \citep{Bentz06a}
which was abbreviated by poor weather.  The other three AGNs ---
NGC~3227, NGC~3516, and NGC~5548 --- have been targets in multiple
previous reverberation campaigns, and were specifically included in
this campaign to compare the velocity-dependent lags, which might be
interpreted as indicators of the gross kinematics of the BLR, with
previous results from our 2007 campaign \citep{Denney09a}. In all
three cases, the pattern of the lags as a function of
velocity has changed. The most likely reason for this is that
the BLR structure is probably complex and consists of multiple
components --- a disk and a wind, for example
\citep[e.g.,][]{Storchi-Bergmann17} --- and characterizing any
particular velocity bin by a single lag is simply inadequate to
describe the BLR structure and kinematics. The important point is that
the apparent differences between the 2007 and 2012 results
suggest that changes may occur over a BLR dynamical timescale.
In a future contribution, we will undertake a more detailed analysis of these data
with the aim of determining the structure and kinematics of the BLR in these sources 
and determine whether or not the apparent changes are 
significant.

\acknowledgments GDR, CJG, BMP, and RWP are grateful for the support
of the National Science Foundation through grant AST-1008882 to The
Ohio State University. KDD, BJS, CBH, and JLV acknowledge support by
NSF Fellowships. MCB gratefully acknowledges support from the NSF through CAREER grant AST-1253702.
AMM and DMS acknowledge the support of NSF grants
AST-1004756 and AST-1009756. C.S.K. is supported by NSF grant
AST-1515876.  SK is supported at the Technion by the Kitzman
Fellowship and by a grant from the Israel-Niedersachsen collaboration
program. SR is supported at Technion by the Zeff Fellowship. SGS
acknowledges the support to CrAO in the frame of the
`CosmoMicroPhysics' Target Scientific Research Complex Programme of
the National Academy of Sciences of Ukraine (2007-2012). 
MV gratefully acknowledges support from the Danish Council 
for Independent Research via grant no. DFF 4002-00275. VTD
acknowledges the support of the Russian Foundation of Research (RF
project no.\ 12-02-01237-a). The CrAO CCD cameras were purchased
through the US Civilian Research and Development for Independent
States of the Former Soviet Union (CRDF) awards UP1-2116 and
UP1-2549-CR-03. This research has been partly supported by the
Grant-in-Aids of Scientific Research (17104002, 20041003, 21018003,
21018005, 22253002, and 22540247) of the Ministry of Education,
Science, Culture and Sports of Japan. This research has made use of
the NASA/IPAC Extragalactic Database (NED), which is operated by the
Jet Propulsion Laboratory, California Institute of Technology, under
contract with the National Aeronautics and Space Administration.

\software{IRAF \citep{Tody86, Tody93},
LA Cosmic \citep{vanDokkum01},
mapspec \citep{Fausnaugh17a},
ISIS \citep{Alard98, Alard00},
Sexterp \citep{Siverd12},
JAVELIN \citep{Zu11}.
}




\floattable
\input{target_properties}
\floattable
\input{oiiitab}
\floattable
\input{linewindows}
\floattable
\input{contwindows}
\floattable
\input{contall}
\floattable
\input{lineall}
\floattable
\input{lcstats}
\floattable
\input{hbetalags}
\floattable
\input{linewidths}
\floattable
\input{mass_table}

\clearpage



 
\end{document}

%% file: target_properties.tex
\begin{deluxetable}{lccccc}[h]
\tablewidth{0pt}
\tablecaption{Source Properties \label{tab:targets}}
\tablehead{\colhead{Object} & \colhead{$z$} & \colhead{$D_L$} & 
\colhead{$\log \lambda L_{\rm 5100\,\AA}$} 
& \colhead{$\log \lambda L_{\rm host}$} & \colhead{$E(\bv)$}\\
& & \colhead{(Mpc)} &\colhead{(erg\,s$^{-1}$)} &\colhead{(erg\,s$^{-1}$)}
& \colhead{(mag)}\\
\colhead{(1)}&\colhead{(2)}&\colhead{(3)}&\colhead{(4)}&\colhead{(5)}&\colhead{(6)}
 }
\startdata
Mrk 704 & 0.0292 & 128.0 & 43.72 & 43.27 & 0.03 \\
NGC 3227 & 0.0038 & 23.5 & 42.74 & 42.48 & 0.02 \\
NGC 3516 & 0.0088 & 38.1 & 43.29 & 43.21 & 0.04 \\
NGC 4151 & 0.0033 & 13.9 & 42.61 & 42.37 & 0.02 \\
NGC 5548 & 0.0171 & 74.5 & 43.45 & 43.20 & 0.02 \\
\enddata
\tablecomments{Column 2 is taken from the NASA Extragalactic Database.
  Column 3 gives the luminosity distance in a consensus cosmology,
  except for NGC\,3227 and NGC\,4151 as explained in the text (see \S \ref{section:targets}).  
  Column 4 gives the observed luminosity (corrected for Galactic extinction),
  calculated from the observed 5100\,\AA\ rest-frame light curve and
  Column 3 and is corrected for the starlight contribution which is
  given in Column 5.  Column 6 gives the Galactic reddening value from
  \cite{Schlafly11}.}
\end{deluxetable}

%% file: oiiitab.tex
\begin{deluxetable}{lccc}[h]
\tablewidth{0pt}
\tablecaption{[O\,{\sc iii}]\,$\lambda 5007$ Flux Calibration\label{tab:o3tab}}
\tablehead{\colhead{Object} &  \colhead{No.} & 
\colhead{$F$([O{\sc iii}]$\lambda5007$)} & \colhead{Percent}\\
& \colhead{Photometric} & \colhead{($10^{-13}$\,erg\,s$^{-1}$ cm$^{-2}$)}
&\colhead{scatter}\\
\colhead{(1)}&\colhead{(2)}&\colhead{(3)}&\colhead{(4)}
 }
\startdata
Mrk 704  &  21 & $1.31\pm 0.03$   & 0.40 \\
NGC 3227 &  24 & $7.81\pm 0.16$   & 0.19 \\
NGC 3516 &  21 & $4.58\pm 0.07$   & 0.27 \\
NGC 4151 &  20 & $107\pm 2$ & 0.15 \\
NGC 5548 &  21 & $4.91\pm  0.08$   & 0.18  \\
\enddata
\tablecomments{\label{tab:targets}
Column 2 gives the number of nights with clear and stable conditions 
and judged to be photometric.
Each object had three observations per
night, which were used to calculate the narrow 
[O{\sc iii}]$\lambda$5007 line flux.  The line flux and its uncertainty
  are given in Column 3.  Column 4 gives the fractional variation of
  the [O{\sc iii}]$\lambda$5007 line light curve, which serves as an
  estimate of the night-to-night calibration error.}
\end{deluxetable}

%% file: linewindows.tex
\begin{deluxetable}{lcccc}[h]
\tablewidth{0pt}
\tablecaption{Observed-Frame Integration Windows \label{tab:windows}}
\tablehead{\colhead{Object} & \colhead{5100\,\AA} & \colhead{H$\beta$} & \colhead{[O{\sc iii}]\,$\lambda5007$} \\
&\colhead{(\AA)} &\colhead{(\AA)} &\colhead{(\AA)} 
}
\startdata
Mrk 704 & 5250--5270 & 4910--5122 &  5138--5168  \\
NGC 3227 & 5110--5130 & 4812--4942 &  5005--5047  \\
NGC 3516 & 5128--5170 & 4775--4960 &  5032--5066  \\
NGC 4151 & 5110--5140 & 4815--4955 &  4998--5055  \\
NGC 5548 & 5179--5210 & 4830--5052 &  5070--5110  \\
\enddata
\end{deluxetable}

%% file: contwindows.tex
\begin{deluxetable}{llccc}[h]
\tablewidth{0pt}
\tablecaption{Observed-Frame Continuum Fitting Windows \label{tab:contwindows}}
\tablehead{\colhead{Object} & \colhead{Line Side} & \colhead{H$\beta$} & 
\colhead{[O{\sc iii}]\,$\lambda4959$} &\colhead{[O{\sc iii}]\,$\lambda5007$ }  \\
& &\colhead{(\AA)}&\colhead{(\AA)}&\colhead{(\AA)}
}
\startdata
Mrk 704 & Blue & 4890--4910 & 5080--5090  & 5128--5137  \\
              & Red & 5235--5245 &  5113--5120 & 5169--5175  \\
NGC 3227 & Blue & 4801--4811 &4950--4961 & 5000--5005   \\
                  & Red & 4941--4950 & 4992--5002& 5047--5055   \\
NGC 3516 & Blue & 4743--4752 & 4970--4980 & 5022--5032   \\
                  & Red & 5128--5170 & 5014--5030 & 5067--5075   \\
NGC 4151 & Blue & 4510--4520 & 4950--4955 & 4993--4998   \\
                  & Red & 5110--5140 & 4990--5000 & 5055--5070   \\
NGC 5548 & Blue & 4535--4545 & 5017--5027& 5060--5070   \\
                  & Red & 5136--5159 & 5058--5070 & 5110--5120   \\
\enddata
\end{deluxetable}

%% file: contall.tex
\rotate
\begin{deluxetable}{cccccccccccc}
\tablewidth{0pt}
\tablecaption{Continuum Light Curves}
\tablehead{
\multicolumn{2}{c}{Mrk 704} &
\multicolumn{2}{c}{NGC 3227} &
\multicolumn{2}{c}{NGC 3227 (2014)} &
\multicolumn{2}{c}{NGC 3516} &
\multicolumn{2}{c}{NGC 4151} &
\multicolumn{2}{c}{NGC 5548} 
\\
\colhead{HJD\tablenotemark{a}} & \colhead{$F_{\lambda}$\tablenotemark{b}} &
\colhead{HJD\tablenotemark{a}} & \colhead{$F_{\lambda}$\tablenotemark{b}} &
\colhead{HJD\tablenotemark{a}} & \colhead{$F_{\lambda}$\tablenotemark{b}} &
\colhead{HJD\tablenotemark{a}} & \colhead{$F_{\lambda}$\tablenotemark{b}} &
\colhead{HJD\tablenotemark{a}} & \colhead{$F_{\lambda}$\tablenotemark{b}} &
\colhead{HJD\tablenotemark{a}} & \colhead{$F_{\lambda}$\tablenotemark{b}} 
}
\startdata
5932.26 & $4.29 \pm 0.05$ M & 5933.82 & $14.74 \pm 0.22$ M & 6645.61 & $10.81 \pm 0.22$ W1 & 5932.39 & $19.16 \pm 0.20$ M & 5931.51 & $35.52 \pm 1.03$ C & 5931.55 & $8.46 \pm 0.22$ C \\
5933.57 & $4.27 \pm 0.02$ W1 & 5935.32 & $15.25 \pm 0.32$ M & 6646.60 & $10.89 \pm 0.23$ W1 & 5933.59 & $18.57 \pm 0.05$ W1 & 5932.38 & $37.03 \pm 0.55$ M & 5932.44 & $8.83 \pm 0.15$ M \\
5933.75 & $4.26 \pm 0.11$ M & 5935.55 & $15.67 \pm 0.02$ W1 & 6647.62 & $11.16 \pm 0.20$ W1 & 5933.86 & $18.71 \pm 0.29$ M & 5933.91 & $36.82 \pm 0.42$ M & 5933.62 & $8.60 \pm 0.05$ W1 \\
5935.26 & $4.30 \pm 0.08$ M & 5936.81 & $15.08 \pm 0.26$ M & 6648.59 & $10.98 \pm 0.14$ W1 & 5934.89 & $19.15 \pm 0.28$ M & 5935.38 & $34.92 \pm 0.30$ M & 5933.94 & $8.50 \pm 0.10$ M \\
5939.50 & $4.44 \pm 0.02$ W1 & 5937.60 & $16.01 \pm 0.15$ A1 & 6650.51 & $11.06 \pm 0.16$ W1 & 5935.85 & $18.43 \pm 0.31$ M & 5935.90 & $34.44 \pm 1.32$ M & 5935.62 & $8.81 \pm 0.03$ W1 \\
5940.75 & $4.32 \pm 0.06$ M & 5937.80 & $14.34 \pm 0.63$ M & 6653.59 & $12.52 \pm 0.34$ W1 & 5936.47 & $19.03 \pm 0.04$ W1 & 5936.51 & $34.22 \pm 0.19$ W1 & 5935.93 & $8.57 \pm 0.11$ M \\
5942.35 & $4.30 \pm 0.03$ W1 & 5938.80 & $15.46 \pm 0.47$ M & 6655.50 & $13.01 \pm 0.37$ W1 & 5936.86 & $18.68 \pm 0.31$ M & 5937.40 & $33.10 \pm 0.26$ M & 5936.54 & $8.75 \pm 0.03$ W1 \\
5943.56 & $4.41 \pm 0.02$ W1 & 5940.55 & $15.64 \pm 0.14$ A1 & 6656.51 & $13.06 \pm 0.37$ W1 & 5937.85 & $18.30 \pm 0.31$ M & 5937.56 & $33.65 \pm 1.33$ A1 & 5936.93 & $8.55 \pm 0.12$ M \\
5944.76 & $4.27 \pm 0.06$ M & 5940.80 & $15.60 \pm 0.25$ M & 6661.93 & $13.70 \pm 0.16$ M & 5938.14 & $20.18 \pm 0.08$ A2 & 5938.65 & $33.64 \pm 1.33$ A1 & 5937.93 & $8.49 \pm 0.11$ M \\
5945.74 & $4.30 \pm 0.06$ M & 5941.48 & $15.65 \pm 0.14$ A1 & 6662.48 & $13.07 \pm 0.26$ W1 & 5939.63 & $19.37 \pm 0.05$ W1 & 5939.62 & $33.30 \pm 1.32$ A1 & 5938.59 & $8.34 \pm 0.24$ A1 \\
5946.75 & $4.34 \pm 0.05$ M & 5942.47 & $15.88 \pm 0.15$ A1 & 6663.50 & $13.00 \pm 0.41$ W1 & 5939.82 & $19.46 \pm 0.32$ M & 5939.65 & $33.52 \pm 0.21$ W1 & 5939.69 & $8.88 \pm 0.26$ A1 \\
5947.75 & $4.30 \pm 0.04$ M & 5942.54 & $15.93 \pm 0.02$ W1 & 6663.90 & $13.75 \pm 0.13$ M & 5942.59 & $19.69 \pm 0.07$ W1 & 5940.59 & $33.46 \pm 1.32$ A1 & 5940.43 & $8.03 \pm 0.05$ M \\
5949.33 & $4.55 \pm 0.10$ M & 5943.45 & $16.01 \pm 0.15$ A1 & 6664.88 & $13.94 \pm 0.14$ M & 5943.60 & $20.06 \pm 0.08$ W1 & 5940.89 & $33.57 \pm 0.35$ M & 5940.61 & $8.06 \pm 0.23$ A1 \\
5951.75 & $4.63 \pm 0.04$ M & 5943.60 & $16.39 \pm 0.03$ W1 & 6665.42 & $14.26 \pm 0.21$ W1 & 5944.87 & $19.93 \pm 0.30$ M & 5941.58 & $33.46 \pm 1.32$ A1 & 5941.60 & $8.00 \pm 0.23$ A1 \\
5952.45 & $4.48 \pm 0.03$ W1 & 5944.01 & $16.14 \pm 0.69$ M & 6666.93 & $14.02 \pm 0.17$ M & 5945.84 & $19.77 \pm 0.32$ M & 5942.57 & $34.30 \pm 1.35$ A1 & 5942.59 & $7.91 \pm 0.23$ A1 \\
5952.76 & $4.79 \pm 0.08$ M & 5944.44 & $15.90 \pm 0.15$ A1 & 6666.94 & $13.95 \pm 0.14$ M & 5946.89 & $19.50 \pm 0.30$ M & 5942.62 & $33.43 \pm 0.23$ W1 & 5942.62 & $7.92 \pm 0.04$ W1 \\
5953.45 & $4.52 \pm 0.03$ W1 & 5945.31 & $16.14 \pm 0.14$ M & 6670.45 & $13.35 \pm 0.17$ W1 & 5947.11 & $19.43 \pm 0.08$ A2 & 5943.56 & $34.61 \pm 1.36$ A1 & 5943.58 & $7.72 \pm 0.22$ A1 \\
5953.75 & $4.46 \pm 0.05$ M & 5945.44 & $15.86 \pm 0.14$ A1 & 6671.63 & $13.66 \pm 0.21$ C & 5948.65 & $20.10 \pm 0.07$ W1 & 5943.62 & $33.74 \pm 0.17$ W1 & 5943.59 & $8.00 \pm 0.04$ W1 \\
5955.40 & $4.51 \pm 0.03$ W1 & 5946.49 & $16.39 \pm 0.16$ A1 & 6671.89 & $12.96 \pm 0.30$ FWO & 5948.94 & $20.31 \pm 0.31$ M & 5944.53 & $32.46 \pm 1.29$ A1 & 5945.43 & $7.95 \pm 0.05$ M \\
5955.75 & $4.54 \pm 0.08$ M & 5947.33 & $15.84 \pm 0.14$ M & 6672.45 & $13.09 \pm 0.17$ W1 & 5949.86 & $20.08 \pm 0.32$ M & 5945.40 & $31.77 \pm 0.48$ M & 5947.45 & $8.04 \pm 0.05$ M \\
5956.24 & $4.57 \pm 0.01$ W1 & 5948.52 & $16.38 \pm 0.16$ A1 & 6672.91 & $13.19 \pm 0.28$ FWO & 5950.63 & $20.60 \pm 0.08$ W1 & 5945.54 & $30.85 \pm 1.25$ A1 & 5948.64 & $8.21 \pm 0.24$ A1 \\
5956.75 & $4.54 \pm 0.05$ M & 5948.65 & $16.36 \pm 0.03$ W1 & 6674.00 & $13.27 \pm 0.27$ FWO & 5951.86 & $20.28 \pm 0.31$ M & 5946.93 & $31.13 \pm 0.38$ M & 5948.65 & $8.26 \pm 0.04$ W1 \\
5957.76 & $4.58 \pm 0.09$ M & 5949.46 & $16.20 \pm 0.15$ A1 & 6679.44 & $13.44 \pm 0.13$ W1 & 5952.61 & $20.33 \pm 0.09$ W1 & 5947.90 & $32.61 \pm 0.71$ M & 5949.94 & $8.14 \pm 0.09$ M \\
5958.60 & $4.57 \pm 0.03$ W1 & 5949.82 & $16.58 \pm 0.28$ M & 6679.91 & $13.83 \pm 0.32$ FWO & 5952.87 & $21.06 \pm 0.33$ M & 5948.61 & $32.28 \pm 1.29$ A1 & 5950.62 & $8.01 \pm 0.23$ A1 \\
5958.77 & $4.52 \pm 0.04$ M & 5950.63 & $16.22 \pm 0.03$ W1 & 6682.40 & $14.58 \pm 0.27$ W1 & 5953.57 & $20.52 \pm 0.10$ W1 & 5948.65 & $32.69 \pm 0.24$ W1 & 5950.64 & $8.11 \pm 0.05$ W1 \\
\dots   &\dots &\dots   &\dots &\dots   &\dots &\dots   &\dots &\dots   &\dots &\dots   &\dots \\
\enddata
\tablenotetext{a}{Days $-$ 2\,450\,000} 
\tablenotetext{a}{$10^{-15}$  \,ergs\,s$^{-1}$\,cm$^{-2}$\,\AA$^{-1}$} 
\tablecomments{The
  alpha-numeric codes in the $F_{\lambda}$ columns indicate the
  contributing observatory: M is MDM, W1 is Wise,
  W2 is West Mountain, A1 is Asiago, A2 is ASAS, C is Crimean
  Astrophysical, and F is Fountainwood.  A machine-readable version of
  this table is published in the electronic edition of this article. A
  portion is shown here for guidance regarding its form and content.}
\end{deluxetable}

%% file: lineall.tex
\rotate
\begin{deluxetable}{cccccccccccc}
\tablewidth{0pt}
\tablecaption{H$\beta$ Light Curves}
\tablehead{
\multicolumn{2}{c}{Mrk 704} &
\multicolumn{2}{c}{NGC 3227} &
\multicolumn{2}{c}{NGC 3227 (2014)} &
\multicolumn{2}{c}{NGC 3516} &
\multicolumn{2}{c}{NGC 4151} &
\multicolumn{2}{c}{NGC 5548} 
\\
\colhead{HJD\tablenotemark{a}} & \colhead{$F$\tablenotemark{b}} &
\colhead{HJD\tablenotemark{a}} & \colhead{$F$\tablenotemark{b}} &
\colhead{HJD\tablenotemark{a}} & \colhead{$F$\tablenotemark{b}} &
\colhead{HJD\tablenotemark{a}} & \colhead{$F$\tablenotemark{b}} &
\colhead{HJD\tablenotemark{a}} & \colhead{$F$\tablenotemark{b}} &
\colhead{HJD\tablenotemark{a}} & \colhead{$F$\tablenotemark{b}} 
}
\startdata
5932.26 & $2.92 \pm 0.07$ M & 5933.82 & $4.90 \pm 0.20$ M & 6661.93 & $3.48 \pm 0.15$ M & 5932.39 & $5.58 \pm 0.08$ M & 5931.51 & $48.24 \pm 1.20$ C & 5931.55 & $7.08 \pm 0.15$ C \\
5933.75 & $2.94 \pm 0.12$ M & 5935.32 & $4.89 \pm 0.08$ M & 6663.90 & $3.61 \pm 0.05$ M & 5933.86 & $5.62 \pm 0.07$ M & 5932.38 & $47.52 \pm 0.55$ M & 5932.44 & $7.19 \pm 0.10$ M \\
5935.26 & $2.93 \pm 0.08$ M & 5936.81 & $4.93 \pm 0.04$ M & 6664.88 & $3.74 \pm 0.06$ M & 5934.89 & $5.37 \pm 0.21$ M & 5933.91 & $48.33 \pm 0.95$ M & 5933.94 & $7.12 \pm 0.05$ M \\
5940.75 & $3.02 \pm 0.05$ M & 5937.60 & $5.38 \pm 0.16$ A1 & 6666.93 & $3.90 \pm 0.09$ M & 5935.85 & $5.55 \pm 0.27$ M & 5935.38 & $46.72 \pm 0.33$ M & 5935.93 & $6.92 \pm 0.10$ M \\
5944.76 & $2.99 \pm 0.03$ M & 5937.80 & $5.05 \pm 0.11$ M & 6666.94 & $3.97 \pm 0.09$ M & 5936.86 & $5.48 \pm 0.12$ M & 5935.90 & $46.44 \pm 0.24$ M & 5936.93 & $6.89 \pm 0.15$ M \\
5945.74 & $2.96 \pm 0.02$ M & 5938.80 & $5.39 \pm 0.55$ M & 6777.72 & $4.30 \pm 0.05$ M & 5937.85 & $5.51 \pm 0.26$ M & 5937.40 & $45.51 \pm 0.42$ M & 5937.93 & $6.89 \pm 0.32$ M \\
5946.75 & $2.96 \pm 0.02$ M & 5940.55 & $4.99 \pm 0.15$ A1 & 6778.70 & $4.13 \pm 0.07$ M & 5939.82 & $5.62 \pm 0.20$ M & 5937.56 & $46.44 \pm 1.12$ A1 & 5938.59 & $6.90 \pm 0.16$ A1 \\
5947.75 & $2.97 \pm 0.02$ M & 5940.80 & $4.95 \pm 0.17$ M & 6779.71 & $4.19 \pm 0.05$ M & 5944.87 & $5.75 \pm 0.07$ M & 5938.65 & $46.52 \pm 1.13$ A1 & 5939.69 & $6.36 \pm 0.14$ A1 \\
5949.33 & $2.88 \pm 0.03$ M & 5941.48 & $5.00 \pm 0.15$ A1 & 6780.71 & $4.14 \pm 0.05$ M & 5945.84 & $5.68 \pm 0.07$ M & 5939.62 & $43.57 \pm 1.04$ A1 & 5940.43 & $6.60 \pm 0.14$ M \\
5951.75 & $2.95 \pm 0.05$ M & 5942.47 & $5.24 \pm 0.15$ A1 & 6781.70 & $4.08 \pm 0.07$ M & 5946.89 & $5.95 \pm 0.11$ M & 5940.59 & $44.43 \pm 1.06$ A1 & 5940.61 & $6.66 \pm 0.15$ A1 \\
5952.76 & $2.89 \pm 0.06$ M & 5943.45 & $5.38 \pm 0.16$ A1 & 6782.71 & $4.04 \pm 0.06$ M & 5948.94 & $5.48 \pm 0.09$ M & 5940.77 & $45.29 \pm 0.74$ M & 5941.60 & $6.52 \pm 0.15$ A1 \\
5953.75 & $2.93 \pm 0.04$ M & 5944.01 & $5.27 \pm 0.19$ M & 6783.70 & $4.00 \pm 0.07$ M & 5949.86 & $5.80 \pm 0.12$ M & 5941.58 & $45.53 \pm 1.10$ A1 & 5942.59 & $6.53 \pm 0.15$ A1 \\
5955.75 & $2.82 \pm 0.05$ M & 5944.44 & $5.26 \pm 0.15$ A1 & 6784.77 & $4.05 \pm 0.09$ M & 5951.86 & $6.09 \pm 0.08$ M & 5942.57 & $45.37 \pm 1.09$ A1 & 5943.58 & $6.49 \pm 0.15$ A1 \\
5956.75 & $3.00 \pm 0.05$ M & 5945.31 & $5.52 \pm 0.06$ M & 6785.70 & $3.86 \pm 0.06$ M & 5952.87 & $6.03 \pm 0.17$ M & 5943.56 & $46.06 \pm 1.11$ A1 & 5945.43 & $6.17 \pm 0.07$ M \\
5957.76 & $2.94 \pm 0.07$ M & 5945.44 & $5.23 \pm 0.15$ A1 & 6786.71 & $3.96 \pm 0.06$ M & 5953.85 & $5.97 \pm 0.16$ M & 5944.53 & $43.51 \pm 1.04$ A1 & 5947.45 & $6.19 \pm 0.05$ M \\
5958.77 & $2.88 \pm 0.03$ M & 5946.49 & $5.79 \pm 0.17$ A1 & 6787.69 & $3.96 \pm 0.06$ M & 5955.01 & $6.41 \pm 0.18$ M & 5945.40 & $44.06 \pm 0.32$ M & 5948.64 & $5.94 \pm 0.13$ A1 \\
5960.27 & $2.97 \pm 0.02$ M & 5947.33 & $5.58 \pm 0.03$ M & 6788.70 & $4.03 \pm 0.07$ M & 5955.86 & $6.13 \pm 0.11$ M & 5945.54 & $43.99 \pm 1.05$ A1 & 5949.94 & $5.86 \pm 0.10$ M \\
5961.75 & $2.97 \pm 0.14$ M & 5948.52 & $5.78 \pm 0.17$ A1 & 6789.71 & $4.06 \pm 0.07$ M & 5956.86 & $6.15 \pm 0.08$ M & 5946.93 & $43.56 \pm 0.29$ M & 5950.62 & $6.14 \pm 0.13$ A1 \\
5962.77 & $2.96 \pm 0.05$ M & 5949.46 & $5.58 \pm 0.16$ A1 & 6790.68 & $4.03 \pm 0.08$ M & 5957.87 & $6.53 \pm 0.11$ M & 5947.90 & $44.77 \pm 1.17$ M & 5951.54 & $5.97 \pm 0.13$ A1 \\
5963.50 & $2.98 \pm 0.08$ M & 5949.82 & $5.57 \pm 0.09$ M & 6791.69 & $3.81 \pm 0.07$ M & 5958.88 & $6.05 \pm 0.23$ M & 5948.61 & $42.95 \pm 1.02$ A1 & 5951.95 & $6.21 \pm 0.08$ M \\
5968.27 & $2.98 \pm 0.07$ M & 5951.45 & $5.62 \pm 0.16$ A1 & 6792.69 & $3.90 \pm 0.07$ M & 5959.88 & $6.30 \pm 0.07$ M & 5949.03 & $43.95 \pm 0.32$ M & 5953.45 & $6.08 \pm 0.07$ M \\
5973.69 & $3.10 \pm 0.05$ M & 5951.80 & $5.75 \pm 0.04$ M & 6793.70 & $3.94 \pm 0.07$ M & 5960.86 & $6.10 \pm 0.09$ M & 5949.55 & $42.68 \pm 1.01$ A1 & 5953.69 & $6.04 \pm 0.13$ A1 \\
5978.76 & $3.16 \pm 0.05$ M & 5952.70 & $5.78 \pm 0.17$ A1 & 6795.69 & $4.00 \pm 0.10$ M & 5961.86 & $6.48 \pm 0.18$ M & 5949.91 & $42.63 \pm 0.23$ M & 5955.52 & $6.28 \pm 0.12$ C \\
5979.76 & $3.26 \pm 0.08$ M & 5953.32 & $5.68 \pm 0.08$ M & 6798.71 & $4.19 \pm 0.06$ M & 5962.88 & $6.17 \pm 0.16$ M & 5950.59 & $44.28 \pm 1.06$ A1 & 5956.54 & $5.99 \pm 0.07$ M \\
5980.73 & $3.24 \pm 0.06$ M & 5955.81 & $5.60 \pm 0.05$ M & 6799.70 & $4.07 \pm 0.06$ M & 5964.24 & $6.07 \pm 0.46$ M & 5951.52 & $43.22 \pm 1.03$ A1 & 5957.92 & $6.20 \pm 0.14$ M \\
\dots   &\dots &\dots   &\dots &\dots   &\dots &\dots   &\dots &\dots   &\dots &\dots   &\dots \\
\enddata
\tablenotetext{a}{Days $-$ 2\,450\,000} 
\tablenotetext{a}{$10^{-13}$  \,ergs\,s$^{-1}$\,cm$^{-2}$} 
\tablecomments{The alpha-numeric codes
  in the $F_{\lambda}$ columns indicate the contributing observatory:
  M is MDM, A1 is Asiago, and  C is Crimean Astrophysical.  A
  machine-readable version of this table is published in the
  electronic edition of this article. A portion is shown here for
  guidance regarding its form and content.}
\end{deluxetable}

%% file: lcstats.tex
\begin{deluxetable}{llcccccccc}
\tablewidth{0pt}
\tablecaption{Light Curve Properties \label{tab:lc_prop}}
\tablehead{
\colhead{Object} & \colhead{Light curve} & \colhead{$N_{\rm obs}$} & \colhead{$\Delta t_{\rm med}$}& \colhead{Uncertainty}  & \colhead{$\langle F \rangle$} & \colhead{$\langle S/N \rangle$} & \colhead{$\sigma_{\rm var}$} & \colhead{$(S/N)_{var}$}&\colhead{$r_{\rm max}$}\\
& & &\colhead{(days)}  &\colhead{Rescaling Factor}& & & & \\
\colhead{(1)}&\colhead{(2)} &\colhead{(3)}&\colhead{(4)}&\colhead{(5)}&\colhead{(6)}&\colhead{(7)}&\colhead{(8)}&\colhead{(9)}&\colhead{(10)}
}
\startdata
Mrk 704 & 5100\,\AA & 97 & 1.01 & 1.53 & 4.81 & 205.5 & 0.06 & 88.1 & \dots \\
 & H$\beta$ & 72 & 1.01 & 1.50 & 3.30 & 89.7 & 0.07 & 40.0 & $ 0.92\pm  0.02$ \\
NGC 3227 & 5100\,\AA & 94 & 1.01 & 2.53 & 15.60 & 175.0 & 0.02 & 27.5 & \dots \\
 & H$\beta$ & 84 & 1.02 & 1.97 & 4.94 & 74.8 & 0.08 & 40.3 & $ 0.69\pm  0.06$ \\
NGC 3227 (2014) & 5100\,\AA & 56 & 1.00 & 1.28 & 13.60 & 99.8 & 0.05 & 24.2 & \dots \\
 & H$\beta$ & 34 & 1.00 & 1.41 & 4.14 & 55.1 & 0.04 & 9.0 & $ 0.77\pm  0.05$ \\
NGC 3516 & 5100\,\AA & 109 & 1.03 & 1.64 & 19.70 & 226.5 & 0.03 & 54.9 & \dots \\
 & H$\beta$ & 77 & 1.01 & 2.54 & 5.96 & 46.8 & 0.07 & 20.1 & $ 0.76\pm  0.04$ \\
NGC 4151 & 5100\,\AA & 119 & 1.01 & 3.09 & 32.20 & 159.3 & 0.11 & 131.7 & \dots \\
 & H$\beta$ & 97 & 1.01 & 3.59 & 40.10 & 111.7 & 0.13 & 103.7 & $ 0.94\pm  0.01$ \\
NGC 5548 & 5100\,\AA & 112 & 1.03 & 1.84 & 7.94 & 156.3 & 0.07 & 87.6 & \dots \\
 & H$\beta$ & 91 & 1.03 & 2.61 & 5.77 & 64.2 & 0.10 & 44.7 & $ 0.74\pm  0.04$ \\
\enddata
\tablecomments{Column 3 gives the number of observations in each light
  curve. Column 4 gives the median cadence.  Column 5 gives the
  rescaling factor by which the statistical uncertainties are
  multiplied to account for additional systematic errors (see
  \S \ref{section:lightcurves}).  Column 6 gives the mean flux level of each
  light curve.  The rest-frame 5100\,\AA\ continuum light curves are
  in units of $10^{-15}$ erg cm$^{-2}$ s$^{-1}$ \AA$^{-1}$, and the
  emission line light curves are in units of $10^{-13}$ erg cm$^{-2}$
  s$^{-1}$.  Column 7 gives the mean signal-to-noise ratio $\langle
  {\rm S/N} \rangle$.  Column 8 gives the rms fractional variability
  defined in Equation \ref{eq:intrinsicRMSspec}.  Column 9 gives the approximate $S/N$ at
  which we detect variability (see \S 2.3.3).  Column 10 gives the
  maximum value of the interpolated cross correlation function (see
  \S \ref{section:timeseries}).}
\end{deluxetable}

%% file: hbetalags.tex
\begin{deluxetable}{lrrr}
\tablewidth{0pt}
\tablecaption{Rest-Frame H$\beta$ Lags \label{tab:Hbetalags}}
\tablehead{\colhead{Object} & \colhead{$\tau_{\rm cent}$} & \colhead{$\tau_{\rm peak}$} & \colhead{$\tau_{\tt JAV}$}\\
&\colhead{(days)}&\colhead{(days)}&\colhead{(days)}\\
\colhead{(1)}&\colhead{(2)}&\colhead{(3)}&\colhead{(4)}
}
\startdata
Mrk 704 & $12.65_{-2.14}^{+1.49}$ & $14.87_{-2.45}^{+5.85}$ & $14.32^{+0.87}_{-1.06}$ \\
NGC 3227  (2012) & $1.29_{-1.27}^{+1.56}$ & $1.74_{-1.59}^{+1.69}$ & $2.29^{+0.23}_{-0.20}$ \\
NGC 3227  (2014) & $2.58_{-1.31}^{+1.20}$ & $2.80_{-1.60}^{+1.0}$ &$\ldots$\\
NGC 3516 & $5.74_{-2.04}^{+2.26}$ & $4.24_{-3.93}^{+2.16}$ & $8.27^{+1.12}_{-0.64}$ \\
NGC 4151 & $6.82_{-0.57}^{+0.48}$ & $6.50_{-1.39}^{+0.99}$ & $6.58^{+0.19}_{-0.22}$ \\
NGC 5548 & $2.83_{-0.96}^{+0.88}$ & $2.66_{-1.55}^{+1.06}$ & $3.66^{+0.53}_{-0.52}$ \\
\enddata
\tablecomments{Column 2 and Column 3 give the centroids and peaks,
  respectively, of the interpolated cross correlation functions
  (ICCFs).  The uncertainties give the central 68\% confidence
  intervals of the cross-correlation centroid distribution \citep{Peterson98}.   Column 4 gives the lag fit by {\tt JAVELIN}.  The
  uncertainties give the central 68\% confidence intervals of the {\tt
    JAVELIN} posterior lag distributions.  All lags are relative to
  the 5100\,\AA\ continuum light curve and corrected to the
  rest frame.}
\end{deluxetable}

%% file: linewidths.tex
\begin{deluxetable}{lcccccc}
\tablewidth{0pt}
\tablecaption{Rest-Frame H$\beta$ Velocity Measurements \label{tab:v_alt}}
\tablehead{
& &\multicolumn{2}{c}{RMS Spectrum}& \multicolumn{2}{c}{Mean Spectrum}\\
\colhead{Object} & \colhead{Line} & \colhead{$\sigma_{\rm line}$} & \colhead{FWHM} & \colhead{$\sigma_{\rm line}$} & \colhead{FWHM}& \colhead{Smoothing Width}\\
& &\colhead{(km s$^{-1}$)}&\colhead{(km s$^{-1}$)}&\colhead{(km s$^{-1}$)}&\colhead{(km s$^{-1}$)}&\colhead{(km s$^{-1}$)}\\
\colhead{(1)}&\colhead{(2)}&\colhead{(3)}&\colhead{(4)}&\colhead{(5)}&\colhead{(6)}&\colhead{(7)}
}
\startdata
Mrk 704 & H$\beta$ & $ 1860_{-130}^{+108}$ & $ 3406_{-240}^{+310}$ & $ 2650_{-3}^{+4}$ & $ 3502_{-30}^{+32}$ & 294 \\
NGC 3227 (2012) & H$\beta$ & $ 1368_{-37}^{+38}$ & $ 3837_{-107}^{+81}$ & $ 1402_{-2}^{+2}$ & $ 1602_{-17}^{+18}$ & 313 \\
NGC 3227 (2014) & H$\beta$ & $ 1428_{-106}^{+97}$ & $ 2236_{-387}^{+487}$ & $ 1301_{-3}^{+4}$ & $ 1324_{-17}^{+17}$ & 226 \\
NGC 3516 & H$\beta$ & $ 2448_{-74}^{+63}$ & $ 3488_{-146}^{+219}$ & $ 2633_{-3}^{+3}$ & $ 3231_{-15}^{+13}$ & 339 \\
NGC 4151 & H$\beta$ & $ 1940_{-22}^{+22}$ & $ 4393_{-110}^{+110}$ & $ 2078_{-2}^{+2}$ & $  5174_{-32}^{+32}$ & 369 \\
NGC 5548 & H$\beta$ & $ 2772_{-34}^{+33}$ & $ 7038_{-110}^{+133}$ & $ 3056_{-4}^{+3}$ & $ 1094_{-9}^{+10}$ & 329 \\
\enddata
\tablecomments{Column 3 and Column 4 give the rms line width and FWHM
  in the rms spectrum.  Column 5 and Column 6 give the same but in the
  mean spectrum. All values are corrected for instrumental broadening
  and the smoothing introduced by the scaling algorithm (see \S2.3.1)---the FWHM of the Gaussian
  smoothing kernel is given in Column 7.  Apart from Column 7, all
  values are reported in the rest frame.}
\end{deluxetable}

%% file: mass_table.tex
\begin{deluxetable}{llrccc}
\tablewidth{0pt}
\tablecaption{Black Hole Masses \label{tab:masses}}
\tablehead{\colhead{Object} & \colhead{$\tau_{\tt JAV}$ (days)} & \colhead{$\sigma_{\rm line}$(km s$^{-1}$)} 
& \colhead{$\log {\rm VP} (\Msun)$ (current)} & \colhead{$\log {\rm VP} (\Msun)$ (previous)}& \colhead{$\log M (\Msun)$}\\
\colhead{(1)}&\colhead{(2)}&\colhead{(3)}&\colhead{(4)}&\colhead{(5)}&\colhead{(6)}
}
\startdata
Mrk 704 & $14.19^{+0.87} _{-0.79}$ & $ 1860^{+108}_{-130}$ & $6.98 \pm 0.06$ & $\ldots$ & $7.63 \pm 0.14$ \\
NGC 3227 (2012) & $2.30^{+0.22} _{-0.20}$ & $ 1368^{+ 38}_{- 37}$ & $5.92 \pm 0.05$ 
&$6.21 \pm 0.04$ & $6.57 \pm 0.13$ \\
NGC 3227 (2014) & $2.6\pm 1.0$ & $ 1428^{+ 97}_{-106}$ & $6.01 \pm 0.19$ &$\ldots$& $6.66 \pm 0.24$ \\
NGC 3516 & $8.11^{+0.75} _{-0.58}$ & $ 2448^{+ 63}_{- 74}$ & $6.99 \pm 0.05$ 
&$6.86\pm0.04$& $7.63 \pm 0.13$ \\
NGC 4151 & $6.59^{+0.19} _{-0.21}$ & $ 1940^{+ 22}_{-  22}$ & $6.68 \pm 0.01$ 
&$ 6.93\pm 0.04$ & $7.33 \pm 0.13$ \\
NGC 5548 & $3.68^{+0.43} _{-0.52}$ & $ 2772^{+ 33}_{- 34}$ & $6.74 \pm 0.06$ &
$7.08\pm0.16$& $7.39 \pm 0.14$ \\
\enddata
\tablecomments{Columns 2 and 3 give the time delays measured by {\tt JAVELIN}
and line dispersion in the RMS spectra. The log of the virial product (Eq. 1)
is in Column 4, and previous determinations of the virial product are in 
Column 5: the previous NGC 3227 and NGC 3516 measurements are from
\cite{Denney10}, NGC 4151 is from \cite{Bentz06a}, and the 
value for NGC 5548 is the mean and standard deviation of 16
reverberation results drawn from the literature.
{\bf Column 6: Black hole mass based on the data from
this campaign and assuming $\langle f \rangle = 4.47 \pm 1.25$
\citep{Woo15}.}}
\end{deluxetable}